\documentclass[prb,aps,twocolumn,showpacs,preprintnumbers,amsmath,amssymb]{revtex4-2}

\usepackage{graphicx}
\usepackage{upgreek}

\usepackage[hidelinks]{hyperref}
\hypersetup{
    colorlinks,
    linkcolor={red!50!black},
    citecolor={blue!50!black},
    urlcolor={blue!80!black}
}
\usepackage{tcolorbox}
\usepackage{nicematrix,tikz}

\begin{document}
\title{Nonperturbative quantum field theory for pseudo-Goldstone modes, slow-Goldstone modes, and their quantum chaos}

\author{ Fadi Sun and Jinwu Ye   }
\affiliation{
School of Sciences, Great Bay University, Dongguan 523000, China;\\
Great Bay Institute for Advanced Study, Great Bay University, Dongguan 523808, China;\\
and
Dongguan Key Laboratory for Quantum Black Hole and Quantum Error Correction, Great Bay University, Dongguan 523000, China}
\date{\today }

\begin{abstract}
In this work, we develop a novel form of non-perturbative theory to 
identify a light pseudo-Goldstone mode with a small mass, 
as well as a new type of Goldstone mode with a tiny slope
(termed the slow-Goldstone mode),
which may not be obtained via traditional perturbative methods.
We demonstrate our formalism in the context of superfluids formed by Rashba spin-orbit coupled spinor bosons in a square lattice weakly interacting with a spin-anisotropic interaction.  
The experimental  detections of these two modes, 
especially their roles leading to the quantum information scramblings 
at a finite temperature are discussed. 
The slow-Goldstone mode is compared with 
the slow light and the soft mode in the Sachdev-Ye-Kitaev models.
This non-perturbative formalism can be widely applied to study other emergent particles in various quantum matter. 
\end{abstract}

\maketitle

%\noindent
\section{INTRODUCTION}

Quantum field theory usually is studied by a perturbative expansion 
in terms of a small parameter such as the fine structure constant in QED,
or quark-gluon coupling in the short distance scale 
due to the asymptotic freedom in QCD \cite{QFTbook}. 
However, when studying some strong coupling problems 
such as the strong CP violation problem, 
conventional perturbation theory fails, 
a new non-perturbative method needs to be developed. 
This fact motivates various techniques 
in quantum field theory and string theory to reach non-perturbative results 
from asymptotic series in some coupling constants with zero radius of convergence  
\cite{resurgentdynamics,resurgentmatrix,resurgenttran}.
Similar difficulties may also appear when studying low energy emergent quantum phenomena in quantum matters. 
Indeed, 
 when evaluating various experimentally measurable quantities in quantum materials, one usually apply some popular perturbative expansion 
 in terms of a small parameter such as $ 1/N $ expansion in quantum magnets \cite{scaling,sachdev} and
 Sachdev-Ye-Kitaev model \cite{SY,kittalk,syk2}, 
 or $1/S $ expansion in quantum magnets \cite{gan}
 or a weak interaction in the superfluids \cite{Fetter}.

In particular, retaining only a few lowest order terms in perturbative expansions 
may fail to capture fundamental physics and generate spurious results 
that violate exact symmetry constraints.
Our work investigates superfluid phases of Rashba spin-orbit-coupled spinor bosons 
in a square lattice system with weak spin-anisotropic onsite interactions. 
This system provides a concrete manifestation of such limitations:
naive perturbative treatments yield unphysical gapless excitations 
where symmetry arguments strictly require gapped modes, 
or produce Goldstone modes with erroneous dispersion relations 
that contradict symmetry-breaking predictions. 
Previous approaches incorporating the order from quantum disorder (OFQD) analysis\cite{gan}
– a mechanism that incorporates quantum fluctuations to lift accidental degeneracies 
-- fails to cure these spurious results.
These drawbacks motivated us to develop a new non-perturbative formalism 
to completely remove these spurious results 
and lead to new important physical results consistent with exact symmetry arguments.
In sharp contrast to the results achieved by the various resurgent method 
in math 
\cite{resurgentdynamics} 
and high-energy physics 
\cite{resurgentmatrix,resurgenttran},
these results  on the emergent phenomena in quantum matters 
can be directly measured in the current experiments of 
cold atoms loaded in optical lattices.

The Rashba or Dresselhaus spin-orbit coupling (SOC) 
%\cite{rashba,ahe,socsemi,ahe2,niu,aherev} 
is ubiquitous in various 2d or layered non-centrosymmetric insulators
\cite{niu,aherev}, 
semi-conductor systems
\cite{rashba,ahe2}, 
metals
\cite{ahe}, 
and superconductors
\cite{socsemi}.
The lattice regularization of a linear combination of the Rashba and Dresselhaus SOC,
$ \alpha k_x \sigma_x + \beta k_y \sigma_y $, 
corresponds to the kinetic energy of the system.
The SOC anisotropy can be adjusted by 
%the strains, the shape of the surface or gate electric fields.
strain, surface geometry, or applied gate electric fields.
In the cold-atom experiments, 
there were experimental advances to generate 2d Rashba SOC 
for the fermionic $^{40}$K gas \cite{expk40,expk40zeeman} 
and the tunable quantum anomalous Hall (QAH) SOC 
for bosonic $^{87}$Rb atoms in a square lattice \cite{2dsocbec}.
More recently, the optical lattice clock scheme has been implemented 
to generate very long lifetime 1d SOC for various atoms
\cite{SD-1,SD-2,clock,clock1,clock2,SDRb,ben}. 
The 2d Rashba SOC with a long enough lifetime in a square lattice 
maybe generated in some near future cold-atom experiments.

A strongly interacting Rashba SOC spinor bosons at integer fillings 
in a square lattice was studied in \cite{rh} in the Mott insulating phase.
In this work, we investigate various spin-bond correlated superfluids 
of the same model in the weak interaction limit at arbitrary fillings.
Focus is placed on the anisotropic SOC $ (\alpha = \pi/2, \beta) $ 
with a spin-anisotropic interaction parameter $ \lambda $. 
Due to the heating issue\cite{2dsocbec}, 
the small $\beta$ regime $(0 < \beta < \beta_c \approx 0.29 \pi)$ 
is experimentally more accessible.
In this regime, the analysis is highly dependent on the parameter $ \lambda $,
leading to three distinct phases:

For $\lambda<1$, 
we identify a classical degenerate family of ground states due to a spurious $ U(1) $ symmetry.
The first-step OFQD analysis \cite{gan} 
which lifts spurious degeneracies through zero-point energy corrections,
identifies the plane-wave with spin polarization along x-direction (PW-X) superfluid (SF) phase 
\cite{notation} as the quantum ground state.
This phase features two (gapless) linear modes:
the superfluid Goldstone mode at the momentum $(0,0)$ due to $ U(1)_c $ symmetry breaking,
and a roton mode at the momentum $ (\pi,0) $ %with a spurious linear dispersion 
due to the spurious $ U(1) $ symmetry \cite{subtract}. 
Applying the second-step OFQD method developed in \cite{gan},
we find that the roton mode acquires a small gap with a gap exponent $\nu=1/2$,
thus transformed into a pseudo-Goldstone mode 
(see Fig.\ref{twoexp}(a)) \cite{inter}.

At $\lambda=1$, the superfluid Goldstone mode remains uncritical,
but the roton mode becomes a spurious quadratic one, 
again due to the spurious $ U(1) $ symmetry.
Un-fortunately, the conventional second-step OFQD method \cite{gan} 
fails to resolve these issues, 
prompting the development of a new approach 
that successfully determines the entire spectrum. 
This new %resurgent 
scheme successfully transforms the roton mode 
%with spurious quadratic  dispersion 
into a SOC Goldstone mode with a linear dispersion,
dictated by  the $ U(1)_\text{soc} $ symmetry breaking at $ \lambda = 1 $. 
A notable feature of this SOC Goldstone mode is 
its near-flat dispersion, 
resulting in slower propagation compared to the superfluid Goldstone mode
-hence the name ``slow-Goldstone mode'' (see Fig.\ref{twoexp}(b)).
As a by-product, our %resurgent 
scheme also 
determines the full spectrum of the pseudo-Goldstone mode for $ \lambda < 1 $, 
beyond its gap at zero momentum reached in \cite{gan}.
We analyze the novel quantum information scrambling due to both modes,
and find that the pseudo-Goldstone mode ($\lambda < 1$) 
leads to exponential suppression of the Lyapunov exponent,
while the slow-Goldstone mode ($\lambda=1$) 
leads to $T^3$ scaling of the Lyapunov exponent; 
see Table \ref{tab}.

For $\lambda>1$, 
a classical analysis identifies the ground state
is a superfluid phase 
with spin polarized along Z-direction 
and exhibiting sign alternation between adjacent sites along the $x$-direction.
This phase is referred to as the Z-x SF phase. 
A conventional perturbative calculation in the Z-x SF phase 
opens a gap to the slow-Goldstone mode at $\lambda=1$ 
with the gap exponent $\nu=1$, 
as shown in Fig.~\ref{twoexp}(c).
Thus, the %resurgent 
mechanism generates distinct critical correlation exponents 
on either side of the quantum critical point, 
driven by roton softening and the near-flat slow-Goldstone mode at $\lambda=1$.
A comparison is drawn between the slow-Goldstone mode and the superfluid Goldstone mode 
regarding their distinct Kosterlitz-Thouless (KT) transitions: 
the former drives a spin-orbit KT transition at $T^{\text{soc}}_{\text{KT}}$, 
while the latter governs a superfluid KT transition at higher $T^{\text{SF}}_{\text{KT}}$. 
Our method may also be extended to explore other emergent particles in many other quantum materials.

The paper is organized as follows. 
In Sec. II, we define the Hamiltonian, 
analyze its exact symmetries.
In Sec. III, we perform a mean-field analysis, 
and identify the relevant spurious symmetries. 
Section IV investigates quantum ground states and emergent Goldstone modes 
across the three regimes ($\lambda < 1$, $\lambda = 1$, $\lambda > 1$). 
Finite-temperature phase transitions, quantum chaos behaviors, 
and experimental detection schemes are presented in Sec. V. 
Section VI provides a comparative analysis: 
contrasting the slow-Goldstone mode with slow-light phenomena 
and soft modes in Sachdev-Ye-Kitaev models. 
Some techinical details are provided in four Appendices.

%\section{The Hamiltonian and its symmetries } 

\section{The interacting SOC Model and Exact Symmetries}

\begin{figure}%[!htbp]
\includegraphics[width=\linewidth]{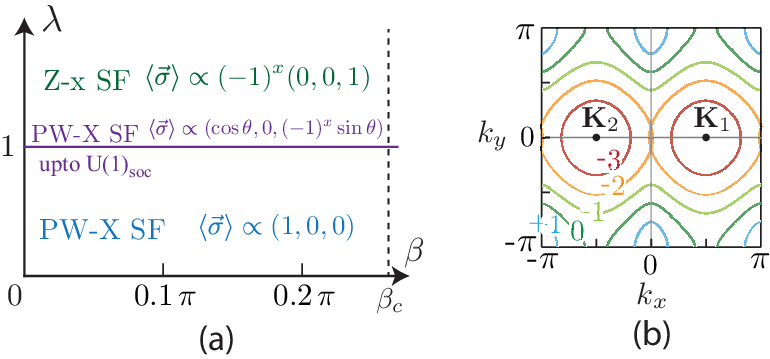}	
\caption{ 
(a) Quantum phase diagram in the $(\beta,\lambda)$ plane 
along the anisotropic SOC line $(\alpha=\pi/2,\beta)$, 
in the weak interaction limit.
The system remains in a superfluid phase throughout.
A $ U(1)_\text{soc} $ symmetry is present at $\lambda=1$,
but is explicitly broken at the spin-dependent interaction $\lambda\neq1$.
(b) Contour plot of the non-interacting spectrum 
$\epsilon_{-}(\mathbf{k})$ for fixed $\beta=\pi/10$. 
There are $ N_K=2 $ minima of $\epsilon_{-}(\mathbf{k})/t$ at $\mathbf{K}_{1,2}=(\pm \pi/2, 0)$ in the experimentally accessible regime $ 0 < \beta < \beta_c \approx 0.29 \pi $
	\cite{NK=4}.
%The physics due to the  $ N_K=4 $ in-commensurate minima at $(\pm \pi/2, \pm k_0 )$, 
%when $ \beta_c  < \beta < \pi/2 $ and the associated quantum Lifshitz transition 
%at $ \beta=\beta_c $ induced by the SF Goldstone boson will be discussed in a separate publication. 
}
\label{sfhalf}
\end{figure}

We study spinor bosons with short-range interactions hopping 
in a square lattice 
subject to a non-Abelian gauge field \cite{rh}:
\begin{align}
	\mathcal{H} 
	= & -t\sum_{\langle i,j\rangle}\sum_{\sigma\sigma'}
	    (b^\dagger_{i\sigma} U_{ij}^{\sigma\sigma'} b_{j\sigma'}+h.c. )   
	    \nonumber  \\
	  & +\frac{U}{2}
	  \sum_{i} (n^2_{i \uparrow} + n^2_{i \downarrow} + 
		    2 \lambda\, n_{i \uparrow} n_{i \downarrow} )
	    - \mu \sum_{i,\sigma} n_{i\sigma} \>,
\label{intlambda}
\end{align}
where $b_{i\sigma}$ ($b_{i\sigma}^\dagger$) annihilates (creates) 
a boson with spin $\sigma=\{\uparrow,\downarrow\}$ at site $i$,
$n_{i\sigma} = b_{i\sigma}^\dagger b_{i\sigma}$ is the number operator,
$\langle i,j\rangle$ denotes nearest-neighbor pairs,
$t$ is the hopping amplitude, 
$U$ ($\lambda U$) parametrizes the intra-spin (inter-spin) repulsive interaction,
and $\mu$ is the chemical potential. 
$U_{i,i+\hat{x}}=e^{i \alpha \sigma_x}$ and $U_{i,i+\hat{y}}=e^{i \beta \sigma_y} $ 
are the non-Abelian gauge potentials put on the two links in the square lattice.
Setting $ \alpha=\beta=0 $, Eq.\eqref{intlambda} reduces to the conventional 
pesudospin-1/2 (two-component) Bose-Hubbard model.
This work focuses on various novel superfluids 
phases arising from anisotropic SOC
($\alpha=\pi/2$ and $ 0 < \beta < \beta_c\approx 0.29 \pi $)
and spin-anisotropic interaction $\lambda$. 
The quantum phase diagram is shown in Fig.\ref{sfhalf}(a).

In the weak interaction limit $U \ll t $, 
one can first diagonalize the hopping term,
and then treat the interaction $ U $ as a small perturbation,
leading to an asymptotic expansion series.
For fixed $\alpha=\pi/2$ and and $0\leq\beta<\beta_c$,
the hopping part in momentum space is:  
$\mathcal{H}_0=-2t\sum_{\mathbf{k}}
b_\mathbf{k}^\dagger 
[\cos\beta\cos k_y-\sin k_x\sigma_x-\sin\beta\sin k_y\sigma_y] 
b_\mathbf{k}$.
The two eigen-energies takes the form:
$\epsilon_{\pm}(\mathbf{k})
=-2t(\cos\beta\cos k_y\mp
\sqrt{\smash[b]{\sin^2 k_x+\sin^2\beta\sin^2 k_y}})$.
As shown in Fig.\ref{sfhalf}(b), 
$\epsilon_{-}(\mathbf{k})$
developes $N_K=2$ minima at $\mathbf{k}=(\pm\pi/2,0)$,
and the corresponding eigenspinors are 
$(1,\mp 1)^\intercal/\sqrt{2}$, respectively.
Note that the eigenspinors are independent of the SOC parameter $\beta$,
which may cause a spurious symmetry 
arise from the higher symmetry case at $\beta=0$.

To analyze the exact symmetries of the model, 
we define the spin operators as
$S_i^{a}=\sum_{\sigma\sigma'}b_{i\sigma}^\dagger\sigma_{\sigma\sigma'}^ab_{i\sigma'}$,
where $a=x,y,z$.
In the special limit $\lambda=1$ and $\beta=0$, %$[U_{i,i+\hat{x}},U_{i,i+\hat{y}}]=0$, 
the non-Abelian gauge field on the $x$-links 
can be gauged away through the transformation
$b_j\to \exp(-i\alpha j_x\sigma_x)b_j$.
This transformation reveals a hidden
spin-orbital coupled SU(2) symmetry generated by:
$Q_X=\sum_iS_i^x$,
$Q_Y=\sum_i(-1)^{i_x}S_i^y$,
and 
$Q_Z=\sum_i(-1)^{i_x}S_i^z$.
Derivations from $\lambda=1$ and $\beta=0$ 
break this symmetry in different ways:
For $\lambda=1$ and $\beta\neq0$,
only $Q_Y$ commutes with the $\mathcal{H}$,
reducing the symmetry to
a spin-orbital coupled $U(1)_\text{soc}$;
for $\lambda\neq1$ and $\beta=0$,
only $Q_Z$ remains conserved;
for $\lambda\neq1$ and $\beta\neq0$,
all three generators are explicitly broken.
In addition to these spin-related symmetries, 
the model always preserves the global charge 
$U(1)_c$ symmetry generated by particle number conservation
$\sum_{i,\sigma} n_{i,\sigma}$.
Furthermore, 
the Hamiltonian $\mathcal{H}$ respects several discrete $ Z_2 $ reflection symmetry:
(1) $ {\cal P}_x $:  $ k_y \rightarrow - k_y, S_i^y \rightarrow -S_i^y,
S_i^x \rightarrow S_i^x,  S_i^z \rightarrow -S_i^z $.
(2) $ {\cal P}_y $ : $ k_x \rightarrow - k_x, S_i^x \rightarrow -S_i^x,
S_i^y \rightarrow S_i^y,  S_i^z \rightarrow -S_i^z $.
(3) $ {\cal P}_z $: $ k_x \rightarrow - k_x, S_i^x \rightarrow -S_i^x,
k_y \rightarrow - k_y, S_i^y \rightarrow -S_i^y, S_i^z \rightarrow S_i^z $.
This $ {\cal P}_z $ symmetry is also equivalent to a joint $ \pi $ rotation of orbital and spin around $ \hat{z} $ axis.

%\section{Mean field Analysis and the origin of the spurious symmetry at $ \lambda \leq 1 $}

\section{Mean-Field Analysis and Origin of Spurious Symmetries}

At $0<\beta<\beta_c$, 
the lower band develops $N_K=2$ minima at 
momenta $(\pm\pi/2,0)$ 
see Fig.\ref{sfhalf}(b).
%and $E_{\rm min}=-2t(1+\cos\beta)$ in Fig.\eqref{sfmott}a1;
%and $E_{\rm min}=-2t\sqrt{1+\csc^2\beta}$ in Fig.\eqref{sfmott}a2.
In the weak interaction regime, most of the bosons condense into 
the single-particle state,
which is superposition of the eigenstates 
at the two degenerate minima of the lower band.
The single-particle wave-function takes the form
\begin{align}
	\Psi_{i}^{0}=\frac{1}{\sqrt{2N_s}}
	\left[
	c_1e^{i\mathbf{K}\cdot\mathbf{r}_i}
	    \begin{pmatrix}
		1\\
		-1\\
	    \end{pmatrix}
	+c_2e^{-i\mathbf{K}\cdot\mathbf{r}_i}
	    \begin{pmatrix}
		1\\
		1\\
	    \end{pmatrix}
	\right]\>,
\label{twonodes}
\end{align}
where $ \mathbf{K}=(\pi/2,0) $,
$N_s$ is the total number of lattice sites,
and  $c_{1,2}$ 
are complex coefficients satisfying the normalization condition 
subject to the normalization condition $|c_1|^2+|c_2|^2=1$.

By construction, the mean-field kinetic energy is independent of $c_{1,2}$. 
The total energy minimization therefore reduces to 
minimizing the interaction energy alone. 
For general $\lambda$, 
the mean-field interaction energy density is:
\begin{align}
    E^0_\text{int}
	=\frac{Un_0^2}{2}\left(1+\frac{\lambda-1}{2}[1-(c_1c_2^*+c_1^*c_2)^2]\right)\>,
\label{E0}
\end{align}
where %$N_s$ is the number of lattice site, 
$N_0$ is the number of condensed atom,
and $n_0=N_0/N_s$ is the condensate density.
Minimization of $E^0_\text{int}$ 
with respect to $c_{1,2}$
yields distinct mean-field ground states. 
%categorized by $\lambda$ values in Fig. \ref{blochsphere}.

\begin{figure}%[!htbp]
\includegraphics[width=\linewidth]{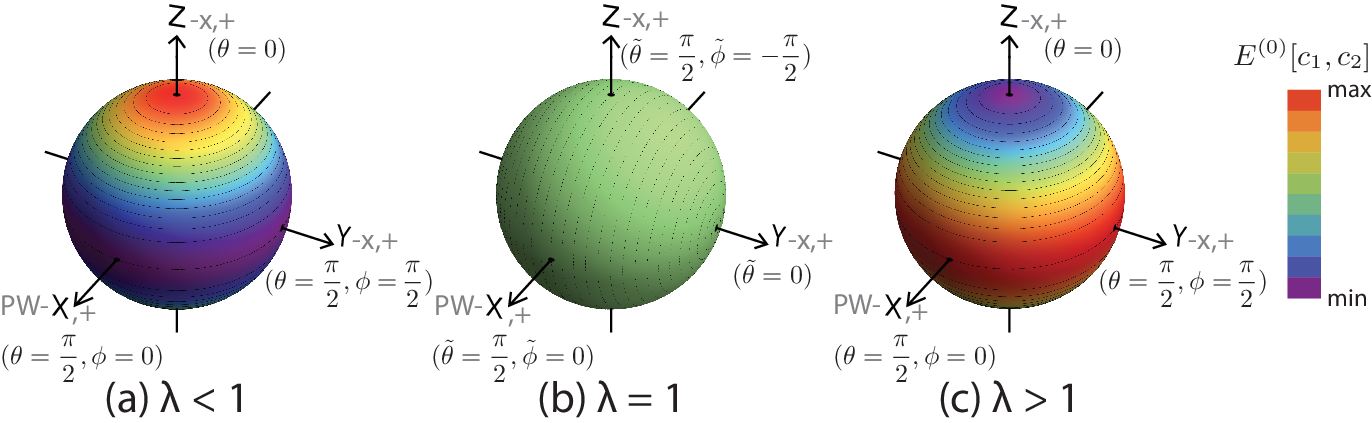}
\caption{ 
Contour plots of the \textsl{classical} ground-state energy $E^{(0)}[c_1,c_2]$ 
on the Bloch sphere %parameter space 
(constrained by $|c_1|^2 + |c_2|^2 = 1$ modulo a global U(1) phase).
%which is a coset $SU(2)/U(1) \simeq S^2$.%forms a 2-sphere
(a) For $\lambda<1$, minima of $E^{(0)}$ occur along the equator.
This spurious degeneracy will be lifted by quantum corrections. 
%The spurious degeneracy at the equator will be lifted by the quantum corrections.
(b) For $\lambda=1$, $E^{(0)}$ remains constant over the entire sphere.
The U(1)$_{\text{soc}}$ symmetry preserves exact degeneracy along meridians 
(dashed lines) 
for arbitrary $\tilde{\phi}$ at fixed $\tilde{\theta}$, 
while quantum corrections will select a specific $\tilde{\theta}$. 
(c) For $\lambda>1$, minima of $E^{(0)}$ localize at north/south pole (Z-x states). }
\label{blochsphere}
\end{figure}

When $\lambda=1$,
the mean-field interaction energy Eq.\eqref{E0}
is independent of $c_{1,2}$.
When $\lambda<1$, 
its minima are reached when $|c_1c_2^*+c_1^*c_2|$ is minimized,
that is $c_1c_2^*+c_1^*c_2=0$.
When $\lambda>1$, 
its minima are reached when $|c_1c_2^*+c_1^*c_2|$ is maximized,
that is $c_1c_2^*+c_1^*c_2=\pm1$.
The naive parameterization
$c_1=e^{+i\phi/2}\cos(\theta/2)$
and $c_2=e^{-i\phi/2}\sin(\theta/2)$
lead to $c_1c_2^*+c_1^*c_2=\cos\phi\sin\theta$.
For $\lambda<1$ this enforces $\cos\phi\sin\theta=0$,
i.e. $\phi=\pm\pi/2$ with arbitray $\theta$.
However, this parameterization becomes singular 
at the plane-wave solution 
— namely $\theta=0$ (where $c_2=0$) and 
$\theta=\pi$ (where $c_1=0$) 
— leaving $\phi$ undefined 
and preventing a unified treatment of the accidental degeneracy. 
To resolve these singularities, 
it is convenient to introduce two Z-x states, 
which satisfy 
$ Q_Z \Psi^0_{\text{Z-x},\pm}= \pm \Psi^0_{\text{Z-x},\pm}$:
\begin{align}
    \Psi^{0}_{\text{Z-x},\pm}=
	\frac{1}{2\sqrt{N_s}}\left[
	    e^{i\mathbf{K}\cdot\mathbf{r}_i}
	    \begin{pmatrix}
		1\\
		-1\\
	    \end{pmatrix}
	    \pm
	    e^{-i\mathbf{K}\cdot\mathbf{r}_i}
	    \begin{pmatrix}
		1\\
		1\\
	    \end{pmatrix}
	\right]\>,
\label{Zxbasis}
\end{align}
which can be called Z-x basis. 
This basis can be used to re-parameterize Eq.\eqref{twonodes} as:
\begin{align}
	\Psi_{i}^{0}=
	e^{i\phi/2}\cos(\theta/2)\Psi^{0}_{\text{Z-x},+}
	+e^{-i\phi/2}\sin(\theta/2)\Psi^{0}_{\text{Z-x},-}\>,
\label{Zx}
\end{align}
which means that in Eq.\eqref{twonodes}:
\begin{align}
    c_1  &=  [e^{i\phi/2}\cos(\theta/2)+e^{-i\phi/2}\sin(\theta/2)]/\sqrt{2} \>,   \nonumber   \\
    c_2  &=  [e^{i\phi/2}\cos(\theta/2)-e^{-i\phi/2}\sin(\theta/2)]/\sqrt{2} \>,
\label{c1c2}
\end{align}
and the interaction energy simplifies to
\begin{align}
    E^{0}_\text{int}=\frac{Un_0^2}{2}
	\left(1+\frac{\lambda-1}{2}\sin^2\theta\right)\>,
\label{classic}
\end{align}
which is explicitly independent of the angle $\phi$ 
%due to the spurious $U(1)$ symmetry 
(see Fig.\ref{blochsphere}).

As discussed in Sec. 3,
turning on a nonzero $\beta$ reduces the exact symmetry:
for $\lambda=1$, $\beta>0$ reduces the $SU(2)$
to exact $U(1)_\text{soc}$ ($Q_Y$ conserved);
for $\lambda\neq1$, $\beta>0$ 
explicitly breaks $Q_Y$ conservation,
leaving no continuous spin symmetry.
Nonetheless, 
since Eq.\eqref{E0} is independent of $\beta$,
the classical ground‑state manifold may exhibit 
an acacidental degeneracy arising from spurious symmetry.
Specifically,
for $\lambda=1$ and $\beta>0$,
the classic ground-state manifold is the (two dimensional) sphere ,
whereas the exact symmetry U(1)$_\mathrm{soc}$ is only one-dimensional.  
This extra degeneracy stems from a spurious U(1) symmetry 
inherited from the original SU(2) at $\beta=0$.
When $\lambda<1$ and $\beta>0$, 
the classic ground-state manifold is the (one dimensional) equator of sphere.
Since there is no exact contiuous spin symmetry,
the continuous degeneracy is due to a spurious $U(1)$ symmetry
inherited from the exact $U(1)$ symmetry at $\beta=0$.
When $\lambda>1$,
the classic ground-state manifold is north/south pole of sphere,
thus the ground-state degeneracy is due to 
spontaneous breaking of the exact symmetry $\mathcal{P}_z$
and no spurious $ U(1) $ symmetry anymore.

%\noindent
%{\bf OFQD on the PW-X SF at $ \lambda < 1 $:    Pseudo-Goldstone (roton ) mode generated by the OFQD:  }
\section{Quantum ground state and various Goldstone modes}

We have determined the classical ground state,
which exhibits accidental degeneracy due to spurious symmetry. 
As will be demonstrated, 
this spurious symmetry generates unphysical artifacts
— not only in ground-state degeneracy but also in low-energy excitations. 
These spurious results will be cured through systematic inclusion of quantum fluctuations within the quantum order from disorder (OFQD) framework. 
The OFQD procedure comprises two steps:
the first step is correcting the ground state energy
and identifying the quantum ground state;
the second step is correcting the low-energy excitations
and obtaining Goldstone modes respecting exact symmetry constraints.

To begin our analysis,
we first review the standard perturbative method for 
bosonic atoms in optical lattices
\cite{Stoof2001}.
In the weak interaction regime,
the spinor field can be decomposed 
into a condensate and fluctuation components:
$ %\Psi=
%\Big(\!\begin{smallmatrix}
%	b_{i\uparrow}\\
%	b_{i\downarrow}\\
%\end{smallmatrix}\!\Big)
b
=\sqrt{N_0} \Psi_0+ \psi $, 
where $N_0$ is the number of condensate atoms
and $\psi$ are bosonic quantum fluctuation fields.
Substituting this decomposition into the Hamiltonian 
yields an asymptotic expansion:
\begin{align}
	\mathcal{H}=\mathcal{H}^{(0)}+\mathcal{H}^{(1)}+\mathcal{H}^{(2)}+\cdots\>, 
%	=1+\frac{\lambda-1}{2}[1-(\delta\theta)^2]
\label{Hasym}
\end{align}
where the superscript denotes the order in the fluctuations $ \psi $ 
and $\cdots$ means higher-order terms.
The leading term $\mathcal{H}^{(0)}=E^{(0)}$ 
gives the classical ground-state energy. 
Setting $\mathcal{H}^{(1)}=0$ determines the chemical potential.
Diagonalizing $ \mathcal{H}^{(2)} $ 
through a Bogoliubov transformation 
$(\psi,\psi^\dagger) \to (\beta,\beta^\dagger)$  
that preserves canonical bosonic commutation relations yields:
\begin{align}
    \mathcal{H}^{(2)}=E^{(2)}_{0}
	+\sum_{\mathbf{k}}
	 \sum_{l}\omega_l(\mathbf{k})
	 \left(\beta_{l,\mathbf{k}}^\dagger\beta_{l,\mathbf{k}}+\frac{1}{2}\right)\>,
\label{h2}
\end{align}
where $\omega_l(\mathbf{k})$ are Bogoliubov excitation energy bands.
These represent the dispersion relations of collective modes in the system, 
with $l$ indexing distinct excitation branches.
The zero-point energy contribution yields
the quantum corrected ground-state energy:
\begin{align}
    E_\text{GS}= E_{0t}
	+\frac{1}{2}\sum_{\mathbf{k}}\sum_{l}\omega_l(\mathbf{k})\>,
\label{h2ground}
\end{align}
with $E_{0t}=E^{(0)} + E^{(2)}_{0}$. 
While this approach is valid without spurious symmetries, 
special care is required when accidental degeneracies exist.
In the following sections, 
we study $\lambda<1$, $\lambda=1$, and $\lambda>1$ in detail.

\begin{figure}%[!htbp]
\includegraphics[width=\linewidth]{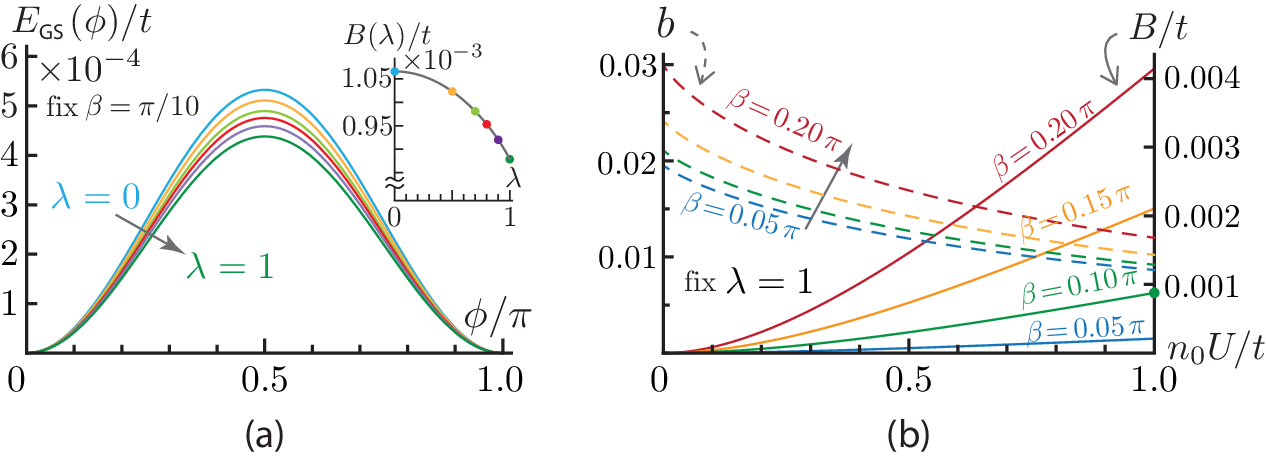}
\caption{ %The quantum $ B $ term generated by the OFQD analysis.
(a) The \textsl{quantum} corrected ground state energy as a function of 
$\phi$ at fixed $\beta=\pi/10$ and $n_0U/t=1$,
for various $\lambda=0,0.5,0.7,0.8,0.9,1.0$.
The $\phi$ variation corresponds to equatorial trajectories 
on the Bloch sphere in Fig.~\ref{blochsphere}.
The insert shows the extracted value for $B$ as a function of $\lambda$.
(b) $B(\lambda=1)$ and $b=Bt/(n_0^2U^2\sin^2\beta)$ 
as a function of $n_0U/t$ for various $\beta=0.05\pi,0.10\pi,0.15\pi,0.20\pi$.
} 
\label{Bterm1}
\end{figure}

\subsection{PW-X SF at $ \lambda < 1 $ and pseudo-Goldstone mode}

For $\lambda < 1$, minimization of the classical ground-state energy 
Eq.\eqref{classic} 
fixes $\theta = \pi/2$ but leaves $\phi$ undetermined. 
This results in a large number of acacidental degenrated ground-state 
due to the spurious $U(1)$ symmetry. 
To resolve this degeneracy, 
we perform an OFQD analysis 
to study quantum corrections to the ground-state energy.

Setting $\theta=\pi/2$ in $\Psi_i^0$ yields 
a wave-function that depends on $\phi$,
denoted $\Psi_i^0(\phi)$.
Decomposing the spinor field as $b_i=\sqrt{N_0}\Psi_i^0(\phi)+\psi$,
generates the asymptotic expansion
$\mathcal{H}=\mathcal{H}^{(0)}
+\mathcal{H}^{(1)}+\mathcal{H}^{(2)}(\phi)+\cdots$,
where 
the classic ground-state energy $\mathcal{H}^{(0)}$ is $\phi$-independent,
$\mathcal{H}^{(1)}=0 $ determines the value of the chemical potential
$ \mu=-2t(1+\cos\beta)+ Un_0(1+\lambda)/2 $,
and $\mathcal{H}^{(2)}$ exhibits explicit $\phi$-dependence.
Diagonalizing $ \mathcal{H}^{(2)} $ through an $ 8 \times 8 $ Bogoliubov transformation 
yields four $\phi$-dependence Bogoliubov bands $\omega_l(\mathbf{k};\phi)$ ($l=1,2,3,4$).
Consequently,
the quantum corrected ground-state energy become $\phi$-dependent.
Numerical evaluation of $E_\text{GS}(\phi)$ via Eq.~\eqref{h2ground} 
in Fig.~\ref{Bterm1} reveals minima at $\phi=0,\pi$.
So the ``quantum order from disorder'' mechanism
selects $\phi=0$ corresponds $(c_1,c_2)=(1,0)$,
or $\phi=\pi$ corresponds $(c_1,c_2)=(0,1)$,
among the classical degenerate family tuned by $ \phi $.
Both configurations correspond to 
spin polarization along the $X$-direction, 
establishing the plane-wave-X superfluid (PW-X SF) phase. 
Besides, these states exhibit opposite spin orientations along $X$ 
and spontaneously break the discrete ${\cal P}_y$ and ${\cal P}_z$ symmetries.

%{\sl (1)  The Excitation spectrum of the PW-X state: the SF Goldstone mode and a spurious linear mode }
\subsubsection{%Excitation Spectrum of the PW-X SF State: 
Superfluid Goldstone Mode and Spurious Linear Mode}

Having established the PW-X SF state as the quantum ground state, 
we now analyze its excitation spectrum and critical behavior 
in the limits $\lambda \to 1^{-}$ and $\beta \to 0^{+}$.

Examining the low-energy properties of 
the lowest Bogoliubov band $\omega_1(k)$,
we find two gapless linear  modes located at 
momentua $(0,0)$ and $(\pi,0)$, respectively.
Expansion around these momenta gives
\begin{align}
    \omega_G(\mathbf{q}) &=\sqrt{n_0 t U(1+\lambda)[ q^2_x+ u(\beta)q_y^2] } \>,\nonumber  \\
    \omega_R(\mathbf{q}) &=\sqrt{n_0 t U(1-\lambda)[ q^2_x+ u(\beta)q_y^2] } \>,
\label{GR}
\end{align}
where $u(\beta)= \cos\beta-\sin^2\beta$,
with momentum measured relative to $(0,0)$ and $(\pi,0)$.
The first mode, $\omega_G$, is the expected superfluid 
Goldstone mode resulting from spontaneous $U_c(1)$ symmetry breaking.
The second mode, $\omega_R$, is the roton mode,
exhibits spurious linear dispersion.
As we demonstrate below, $\omega_R$ will
acquire a small gap through the OFQD mechanism,
and transformed into a pseudo-Goldstone mode.

In the following, 
we will perform the second-step OFQD analysis 
first developed in \cite{gan} on the roton mode.

%{\sl (2) The roton mass gap (pseudo-Goldstone mode) generated from the OFQD mechanism in the PW-X state }

\subsubsection{%Roton Mass Gap in PW-X SF State: 
Pseudo-Goldstone Mode from OFQD Mechanism}

Combining the classical energy
Eq.\eqref{classic} and 
quantum-corrected energy Eq.\eqref{h2ground},
we can extract the $\theta$ and $\phi$ dependence of 
the ground-state energy around its minimum:
\begin{align}
	\mathcal{H}=E_{0t}+\frac{1}{2}A(\delta\theta)^2+\frac{1}{2}B(\delta\phi)^2 \>,
\label{A2B2}
\end{align}
where $\theta=\pi/2+\delta\theta$ and $\phi=0+\delta\phi$.
The coefficient 
$A=\partial^2 E_{int}^0/\partial \theta^2|_{\theta=\pi/2}
=(1-\lambda)Un_0^2/2$ originates from 
the classical contribution,
which is extracted from Eq.\eqref{classic},
while  
$B(\lambda)=\partial^2 E_{GS}/\partial \phi^2|_{\phi=0} 
=b (n_0 U \sin\beta)^2/t $ 
comes from the quantum contribution 
which can be numerically determined from Fig.\ref{Bterm1}(a).

Since $n_0\delta\theta/2$ and $\delta\phi$ form conjugate variables 
satisfying $[n_0\delta\theta/2, \delta\phi] = i\hbar$, 
the second-step OFQD mechanism generates the roton gap \cite{gan}:
\begin{align}
   \Delta^{-}_R & =2\sqrt{AB}/n_0 \sim n_0U \sqrt{(1 - \lambda ) U/t} \>.
\label{gap}
\end{align}
%which is shown in Fig.\eqref{figgap}b to be a monotonically increasing function of $ \beta $.
Obviously, the non-analytic behaviours in both the SOC parameter $ \lambda $ and the weak interaction $ U $ 
can only be achieved from the second-step non-perturbative OFQD mechanism.

%When $ \beta \to 0^{+} $, the roton mode near $ (\pi,0) $ becomes the Goldstone mode due to the breaking of
%$ \tilde{U}(1)_z $ at $ \beta=0 $.  Indeed,  the roton gap vanishes $ \beta \to 0^{+} $ as shown in Fig.4b.
%This limit will be discussed in detail in Sec.VII.

%In fact, as shown in the appendix, the OFQD mechanism not only leads to the
%gap in Eq.\eqref{gap}, but also modifies the dispersion shown in Eq.\eqref{gapdisp}.

\begin{figure}%[!htbp]
\includegraphics[width=\linewidth]{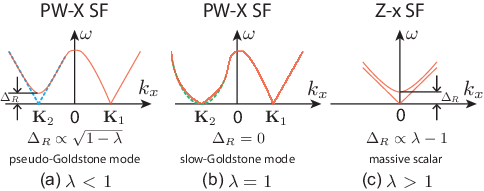}
\caption{ Evolution of the low energy excitations  as $\lambda$ changes from $ \lambda < 1 $ to $ \lambda = 1 $ and $ \lambda > 1 $
at a fixed $ 0 < \beta < \beta_c $.
(a) The tiny roton gap $ \Delta^{-}_R \sim \sqrt{1-\lambda} $ 
generated by non-perturbative quantum corrections at $ \lambda < 1 $.
Its non-analytic behaviour can only be captured by the non-perturbative approach.
The conventional perturbative calculations would lead to the spurious linear mode (dashed line).
(b) The slow-Goldstone mode at $\mathbf{K}_2=(-\pi/2,0) $ due to the $ U(1)_\text{soc} $ breaking generated by non-perturbative quantum corrections at $ \lambda = 1 $. 
Its slope is much smaller than that of the superfluid Goldstone mode at  $\mathbf{K}_1=(\pi/2,0) $.
The conventional perturbative calculations would lead to the spurious quadratic mode (dashed line).
(c) The large roton gap $ \Delta^{+}_R \sim \lambda-1 $ in the Z-x SF state at $ \lambda >1 $. Its analytic behaviour  can be reached simply by a perturbative calculation.
The superfluid Goldstone mode at $ (\pi/2,0) $ from the $ U(1)_c $ breaking remains uncritical through the transition. }
\label{twoexp}
\end{figure}

As demonstrated below, the $B$ term in Eq.\eqref{A2B2}  
remains non-critical as $\lambda$ approaches $ 1^{-} $.
There is a 2nd order quantum phase transition (QPT) from the PW-X to the Z-x phase induced by the roton dropping as $\lambda \to  1^{-} $,
so the quantum critical behaviour of the roton gap $\Delta^{-}_R\propto\sqrt{1-\lambda}$ when 
$\lambda<1$ (See also Fig.\ref{twoexp}(a)).

Unfortunately, the second-step OFQD mechanism developed in \cite{gan} 
can only achieve the roton gap at $ (\pi, 0) $
instead of the whole excitation spectrum. 
This shortcoming can only be overcame by the new %resurgent 
method  developed here. 
The technical details are presented in the two appendices
from both canonical quantization and the path integral approach. 
The final answer is to change the roton mode in Eq.\eqref{GR}  to
\begin{align}
    \omega_R(\mathbf{q})
	= \sqrt{\Delta^2_R + B_R
	 [ q_x^2+(\cos\beta-C\sin\beta^2 ) q_y^2]}\>,
\label{gapdisp}
\end{align}
   where  $\Delta_R=\omega_R(\mathbf{q}=0)=\sqrt{2BU(1-\lambda)}$ is the roton gap in Eq.\eqref{gap}
   generated by the second-step OFQD  mechanism already developed in \cite{gan}. 
   So our unified %resurgent 
   scheme recovers the previously developed OFQD at the zero momentum, 
   but went much beyond.

%{\sl  (3)  A spurious gapless quadratic model as   $ \lambda \to 1^{-} $. }

\subsubsection{Spurious Gapless Quadratic Mode as $\lambda \to 1^{-} $}

  As $ \lambda \to 1^{-} $, the superfluid Goldstone mode remains un-critical, but  the roton mode in Eq.\eqref{GR} becomes a quadratic one:
\begin{equation}
   \omega^0_R(\mathbf{q})=t\sqrt{ [ q^2_x+ u(\beta)q_y^2] [  q_x^2+ v(\beta)q_y^2 ]} \>,
\label{rotonq}
\end{equation}
where $ v(\beta)=\cos\beta-\frac{\sin^2\beta}{1+n_0U/2t} > u(\beta)  $.
  
  This quadratic dispersion is spurious and  can be viewed as the remanent of 
  the ferromagnetic mode due to the $ \tilde{SU}_s(2) $ symmetry breaking at the %left  
  Abelian point $ (\alpha=\pi/2, \beta =0 ) $. 
  In fact, this problem has already been cured in Eq.\eqref{gapdisp};  
  setting $ \lambda=1 $ turns the spurious quadratic mode into
  the linear SOC Goldstone mode in Eq.\eqref{rotonl}.
  In the following section, we take a different basis to show that it will be turned into a linear SOC Goldstone mode with a tiny slope 
  at $ \lambda=1 $  through the newly developed %resurgent 
  mechanism.

%\noindent
%{\bf The resurgent mechanism on the PW-X SF at $  \lambda=1$:  The resurgent SOC   Goldstone mode.  }
\subsection{\bf PW-X SF at $  \lambda=1$ and slow-Goldstone mode}

%\noindent
When $\lambda=1$ and $\beta<\beta_c$, Eq.\eqref{E0} shows that any $c_1$ and $c_2$ give the same interaction energy,
thus they form a classically  degenerate ground-state manifold due to the spurious $ SU(2) $ symmetry originated from the exact
$ \tilde{SU}(2)_s $ symmetry stressed in the Sec.2.

It was known \cite{rh} that the Hamiltonian Eq.\eqref{intlambda} at $\alpha=\pi/2$ and $\lambda=1$
has a spin-orbit coupled $U_\text{soc}(1)$ symmetry $ Q_Y= \sum_i (-1)^{i_x} S^{y}_i $.
In order to distinguish between an accidental degeneracy and the exact degeneracy,
we introduce two Y-x states:
\begin{align}
    \Psi^0_{\text{Y-x},\pm}
    =\frac{1}{2\sqrt{N_s}}\left[
	e^{i\mathbf{K}\cdot\mathbf{r}_i}
	\begin{pmatrix}
	   1\\
	    -1\\
	\end{pmatrix}
	\pm
	ie^{-i\mathbf{K}\cdot\mathbf{r}_i}
	\begin{pmatrix}
	1\\
	1\\
	\end{pmatrix}
	\right]\>,
\label{Yxbasis}
\end{align}
 which satisfy $ Q_Y \Psi^0_{\text{Y-x},\pm}= \pm \Psi^0_{\text{Y-x},\pm} $. 
 It can be called the Y-x basis (see Fig.\ref{blochsphere}(b)) , 
 in contrast to the Z-x  basis (see Fig.\ref{blochsphere}(a)) listed below Eq.\eqref{Zxbasis}. 
 In fact, the two basis are dual to each other under the exchange of variables in Eq.\eqref{exch}.

It can be used to re-parameterize Eq.\eqref{twonodes} as:
\begin{align}
	\Psi_{i}^{0}\!=\!
	e^{i\tilde{\phi}/2}\cos(\tilde{\theta}/2)\Psi^{0}_{\text{Y-x},+}
	\!\!+e^{-i\tilde{\phi}/2}\sin(\tilde{\theta}/2)\Psi^{0}_{\text{Y-x},-} ,
\label{Yx}
\end{align}
which means that in Eq.\eqref{twonodes}:
\begin{align}
    c_1&=[e^{i\tilde{\phi}/2}\cos(\tilde{\theta}/2)
	+e^{-i\tilde{\phi}/2}\sin(\tilde{\theta}/2)]/\sqrt{2}\>,    \nonumber  \\
    c_2&=i[e^{i\tilde{\phi}/2}\cos(\tilde{\theta}/2)
	-e^{-i\tilde{\phi}/2}\sin(\tilde{\theta}/2)]/\sqrt{2}\>,
\label{c1c2i}
\end{align}
  which differs from Eq.\eqref{c1c2} just by an extra ``$i$'' in $ c_2 $ due to the extra ``$i$'' in Eq.\eqref{Yxbasis}.

The $ U(1)_{soc} $ symmetry at $ \lambda=1 $ dictates that the energy should be independent of $ \phi $.
We expect that the classical degeneracy at the angle $ \theta $ will be lifted by quantum effects,
and a unique $\theta$ will be determined by the OFQD mechanism  
(see Fig.\ref{blochsphere}(b)).

 We set $ \phi=0 $ to spontaneously breaks the $ U(1)_{soc} $ symmetry.
 Following the similar procedures as in the Sec. IV A, 
 we obtain Eqs.\eqref{h2},\eqref{h2ground}
 which picks up its minima at $\theta=\pi/2$.
 So the OFQD  mechanism still selects the PW-X state
 $(\theta=\pi/2,\phi=0)$. All the other exactly degenerate states can be generated by changing $ \phi $.
 If we take $ c_1=1, c_2=0 $ PW-X state in Eq.\eqref{twonodes}, 
then the $ U(1)_\text{soc} $ related family is
\begin{align}
    \Psi_i^0 %( \phi )
	\!=\!\frac{1}{\sqrt{2N_s}}\!\left[ \cos\! \frac{\tilde{\phi}}{2}
	e^{i\mathbf{K}\cdot\mathbf{r}_i}\!
	\begin{pmatrix}
	   1\\
	    \!-1\!\\
	\end{pmatrix}
	\!-\sin\! \frac{\tilde{\phi}}{2}
	e^{-i\mathbf{K}\cdot\mathbf{r}_i}\!
	\begin{pmatrix}
	1\\
	1\\
	\end{pmatrix}\!
	\right],
\label{socu1}
\end{align}
whose corresponding spin-orbital structure is 
\begin{align}  
    S_i^x = -\cos \tilde{\phi}, \quad
    S_i^y = 0, \quad
    S_i^z = -(-1)^x \sin \tilde{\phi} \>.
\label{sigmaxyz}
\end{align}  
Setting $\tilde{\phi}=0, \pi$ and $\tilde{\phi}=\pi/2, 3 \pi/2 $ recovers 
the two PW-X states and the Z-x,${\pm}$ states, respectively (see Fig.\ref{blochsphere} ).

By comparing $\lambda<1$ state $\Psi_i^{0}(\theta=\pi/2,\phi)$ in Eq.\eqref{Zx}
with $\lambda=1$ state $\Psi_i^{0}(\tilde{\theta},\tilde{\phi}=0)$ in  Eq.\eqref{Yx}
%( we used the subscript $1$ for $\lambda<1$ and $2$ for $\lambda=1$ case ), 
we can see the relation between the sets of the OFQD variables in the two cases:
\begin{equation}
	\tilde{\theta}=\pi/2-\phi\>,
\label{exch}
\end{equation}
which just shows the $ ( \theta, \phi) $ exchanges in the Z-x basis in Eq.\eqref{Zxbasis} and the Y-x  basis in \eqref{Yxbasis}.
 As claimed below Eq.\eqref{Yxbasis}, the two basis are dual to each other (Fig.\ref{blochsphere} ).

%{\sl (1)  The failure of  the conventional second-step non-perturbative OFQD analysis developed in  \cite{gan} }

\subsubsection{Failure of Conventional OFQD Analysis in \cite{gan}}

 The $U_\text{soc}(1)$ symmetry dictates no $\phi$ dependence, 
 so the ground-state energy an be written as:
\begin{align}
    E_{GS}( \theta ) =E_{GS}( \theta=\pi/2 )+\frac{1}{2}B(\delta\theta)^2
\label{B2}
\end{align}
where $ \delta\theta =\theta-\pi/2 $ and  $B=B(\lambda=1)$ can be extracted
from Fig.\ref{Bterm1}.
The classical $A$ term in Eq.\eqref{A2B2} vanishes as dictated by the exact $U_\text{soc}(1)$ symmetry.
Because $\lim_{\lambda\to1} B(\lambda<1)=B(\lambda=1)$, so $B(\lambda)$ is continuous at $\lambda=1$.
 Unfortunately, plugging $ A =0 $ into Eq,\eqref{gap} still leads to $ \Delta^{-}_R=0 $, so 
 the conventional 2nd-step non-perturbative OFQD analysis developed in \cite {gan} can not lead to any useful information in this case. 
 One must develop a more advanced non-perturbative method which can not only evaluate the gap at $ q=0 $, but also the whole spectrum in the long wavelength limit. 
 This lofty goal was achieved here by developing a %resurgent 
 non-perturbative method . The technical details are presented in the two appendices 
%Method section
from both canonical quantization and the path integral approach.

%{\sl (2) The slow-Goldstone mode generated from 
%the newly developed non-perturbative approach in the PW-X SF state }

\subsubsection{Slow-Goldstone Mode via Newly Developed non-perturbative  Approach}

As shown in the Appendixes A and B, 
the newly developed  %resurgent 
non-perturbative method  successfully 
changes the spurious quadratic dispersion of the roton mode in Eq.\eqref{rotonq} to  a linear one
which is nothing but the %resurgent
SOC Goldstone mode due to the spontaneous breaking of the $U_\text{soc}(1)$ symmetry,
\begin{align}
	\omega_R(q)=\sqrt{\frac{2Bt}{n_0}\left[q_x^2 +  v(\beta)  q_y^2\right]}
\label{rotonl}
\end{align}
where because $B=b(n_0U\sin\beta)^2/t $, so the slope of the %resurgent 
SOC Goldstone mode 
$ v_{soc}  \sim \sqrt{n_0} U $ is much softer than the
   superfluid velocity $ c \sim \sqrt{n_0 U t } $ of the superfluid Goldstone mode in Eq.\eqref{GR}  (see Fig.\ref{twoexp}(b)).
   The ratio $r_G=v_{soc}/c$ between the slope of %resurgent 
   Goldstone mode and superfluid Goldstone mode is ploted in Fig.\ref{rotondrive}b.
   So the %resurgent 
   SOC Goldstone mode travels much slower that of to the superfluid Goldstone mode.
   Due to the wide momentum separation between the superfluid Goldstone and the roton mode, the superfluid Goldstone mode is
   not affected by the  %resurgent 
   non-perturbative analysis.

%\noindent
%{\bf The Z-x SF state at $ \lambda > 1$: Perturbative calculation: }
\subsection{Z-x SF at $ \lambda > 1$ and perturbative calculation }

%\noindent
When $\lambda>1$ and $\beta<\beta_c$, the  minimization of Eq.\eqref{E0} requires $c_1c_2^*+c_1^*c_2=\pm1$
which means $c_1=\pm c_2=e^{i\gamma}/\sqrt{2}$ where $\gamma$ is a globe phase.
So  the two-fold degenerated ground state is  either $\Psi_{\text{Z-x},+}$ with $c_1=c_2=1/\sqrt{2}$  or $\Psi_{\text{Z-x},-}$ with $c_1=-c_2=1/\sqrt{2}$ 
(see Fig.\ref{blochsphere}(c)). 
So the ground state is determined at the mean-field level upto the exact symmetry $ {\cal P}_z $. 

Let us pick up the $\Psi_{\text{Z-x},+}$ state, we find its 4 excitation modes $\omega_{1,2,3,4}(\mathbf{k})$ in ascending order:
$\omega_1(\mathbf{k})$ contains a Goldstone mode at $(0,0)$,
$\omega_2(\mathbf{k})$ contains a roton mode at $(0,0)$ (see Fig.\ref{twoexp}(c))
and $\omega_{3,4}(\mathbf{k})$ are high-energy bands (not shown in Fig.\ref{twoexp}(c)).
 Note that the Z-x SF state breaks the translation symmetry to two sites per unit cell, so
 the Brillioun zone (BZ) in the Z-x state  becomes the Reduced BZ (RBZ)
 with $ -\pi/2 < k_x < \pi/2, -\pi < k_y < \pi $ shown in Fig.\ref{twoexp}(c).

When $\lambda>1$, one can extract the long wave length limit of 
superfluid Goldstone mode $\omega_1$ and the roton mode $\omega_2$:
\begin{align}
    \omega_1(\mathbf{k})
	&=\sqrt{2n_0tU[k_x^2+w(\beta,\lambda)k_y^2]}\>, \nonumber    \\
    \omega_2(\mathbf{k})
	&=(\lambda-1)n_0U+t[k_x^2+x(\beta,\lambda)k_y^2]\>,
\label{lambdalarge}
\end{align}
where 
\begin{align}
	w(\beta,\lambda)
	&=\cos\beta-\frac{\sin^2\beta}{1+(\lambda-1)n_0U/4t}\>,  \nonumber    \\
	x(\beta,\lambda)
	%&=\cos\beta
	%-\frac{4t(4t+\lambda n_0U)\sin^2\beta}
	%{(4t+n_0U)^2-n_0^2U^2(1+(\lambda-1)^2)}\\
	&=\cos\beta
	-\frac{4t(4t+\lambda n_0U)\sin^2\beta}
	{16t^2+8tn_0U-n_0^2U^2(\lambda-1)^2}\>,
\end{align}
and $ \mathbf{k} \in RBZ $. 
Because $ w(\beta, \lambda=1) = u( \beta) $ listed below Eq.\eqref{GR}, 
it recovers the superfluid Goldstone mode in Eq.\eqref{GR} at $ \lambda=1 $.
Note that $ (\lambda-1)U \ll t $, so despite the minus sign in the denominator of  $ x(\beta,\lambda) $, there is no divergency in it.
  
  In contrast to the pseudo-Goldstone mode Eq.\eqref{gapdisp}, the roton mode is non-relativistic with the roton gap: 
\begin{align}
     \Delta^{+}_R=\omega_2(0,0) = (\lambda-1) n_0 U
\label{gap2}
\end{align}  
 which, in contrast to Eq.\eqref{gap}, is analytic in both $ \lambda $ and $ n_0 U $ as shown in Fig.\ref{Bterm1}(c).
 When one gets very close to the QCP such that $ | \lambda-1 | < U/t \ll 1 $, the gap is smaller than that in the Pseudo-Goldstone mode side.
 When one gets far away from  the QCP such that $ | \lambda-1 | > U/t $, the gap is larger than  that in the Pseudo-Goldstone mode side.

% On the other forefront, Spin-orbit coupling (SOC) has played important roles in various condensed matter systems \cite{ahe,socsemi,niu,wu}.
% Recently the investigation and control of SOC \cite{ahe,socsemi,niu,wu} have become subjects of
% intensive research in both condensed matter and cold atom systems after the discovery of the topological insulators \cite{kane,zhang}.
% In the condensed matter side, there are  increasing number of new quantum materials with significant SOC,
% including several new 5d transition metal oxides, multiferroic materials, and heterostructures of transition metal systems \cite{kitpconf}.

\section{Finite temperature behaviours and experimental detections}

%\noindent

\iffalse
 In the previous sections, we explored the pseudo-Goldstone mode and the slow-Goldstone  by the non-perturbative method developed here.
 However, they are just the elementary excitations  of the interacting SOC systems.
 At a finite temperature near the QPT tuned by the spin-anisotropy interaction $ \lambda $ in Fig.\ref{rotondrive}(a), it is important to explore the 
 many-body behaviours of the system which are different, but complementary to the elementary excitations of the system.
 Then an important property of an interacting system at a finite temperature, especially near a QPT is the quantum chaotic behaviours which we will address
 in this section.  We will also explore the intrinsic connections between the quantum chaotic properties of the system and those of the elementary excitations 
 as listed in the Table \ref{tab}.
\fi

In the previous sections,
we examined the pseudo-Goldstone mode and the slow-Goldstone mode 
through the non-perturbative approach developed in this work.
While these modes are only the elementary excitations of the interacting SOC system.
At finite temperature, particularly in the vicinity of the QPT 
driven by the spin-anisotropy interaction $\lambda$, 
as illustrated in Fig.\ref{rotondrive}(a), 
it becomes essential to investigate the many-body behaviors that go beyond, 
yet remain complementary to, the elementary excitations.
A salient feature of interacting systems near a QPT at finite temperature 
is the emergence of quantum chaotic dynamics, 
which will be the focus of this section.
Furthermore, we will elucidate the intrinsic connections 
between the quantum chaotic properties of the system 
and those associated with its elementary excitations, 
as summarized in Table \ref{tab}.

\subsection{Finite temperature phase transitions and  quantum chaos }

\begin{figure}[!b]
\includegraphics[width=\linewidth]{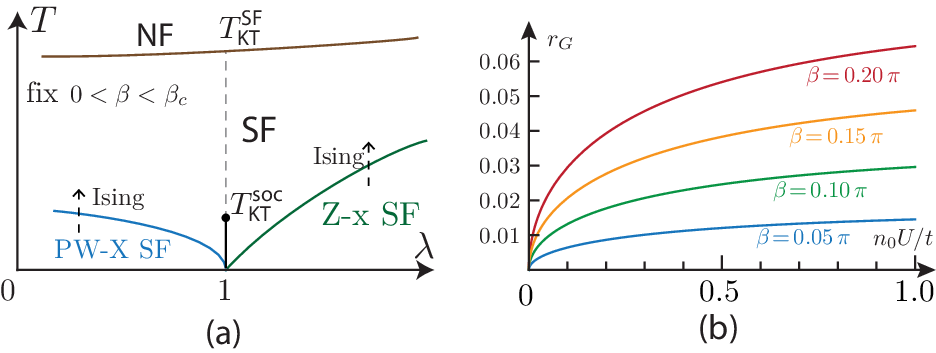}	
\caption{
The finite temperature phase transitions and Quantum chaos driven 
by the roton touchdown corresponding to Fig.\ref{twoexp}.
(a)
%    at a given SOC parameter $ 0 < \beta < \beta_c $, but varying $ \lambda $.
At $ \lambda=1 $, there is a Kosterlitz–Thouless (KT) transition at $ T^\text{soc}_\text{KT} $ due to the slow-Goldstone mode in the SOC sector. 
    There is always a higher KT transition at $ T^\text{SF}_\text{KT} $ from the superfluid to a normal fluid (NF) due to the SF Goldstone mode in the density sector.
	There is a Ising melting transition in the PW-X SF and \mbox{Z-x} SF restoring the discrete $ {\cal P}_y $  symmetry, respectively.  
     The transition temperature $ T_2 \sim \Delta_R \sim |1- \lambda |^{\nu } $ where $  \nu=1/2, 1  $ scales as
      the roton gap at $ (-\pi/2,0) $ in Fig.\ref{twoexp}. At $ \lambda=1 $, the two states become degenerate due to the $ U(1)_\text{soc} $ symmetry.  
    (b) The ratio between the slope of slow-Goldstone mode and superfluid Goldstone mode 
         as a function of $n_0U/t$ at $\lambda=1$ (see Appendix D).   
	 The typical value for $r_G$ is $\sim 1\%$.    }
\label{rotondrive}
\end{figure}

%\noindent
Now we briefly discuss the finite temperature phases and phase transitions above all the SFs in Fig. \ref{sfhalf}.
  Of course, due to the spontaneous $ U(1)_c $ symmetry breaking in all the superfluid phases at $ T=0 $, there is always a KT transition $ T^{SF}_{KT} $ above all the superfluids.  
  At $ \lambda=1 $, due to the  spontaneous $ U(1)_{soc} $ symmetry breaking at $ T=0 $, there is also a KT transition $ T^{soc}_{KT}  $ in Figs.\ref{rotondrive}. 
   Due to the correlated spin-bond orders of the SFs, there are also other interesting phase transitions associated  
   with the restorations of these spin-bond correlated orders at a finite temperature shown in Fig.\ref{rotondrive}. 
   The nature of these finite temperature transitions can be qualitatively determined by the degeneracy of the ground states:
   PW-X ($ d=2 $), Z-x SF ($ d=2 $) which breaks $ {\cal P}_y $ (also $ {\cal P}_z $) and $ {\cal P}_y $ (also $ {\cal P}_x $), respectively.

%In a future publication, we will study the transitions from the various SFs in Fig.1b at weak coupling to the
%magnetic phases at the strong coupling  \cite{rh}.
%At $ \lambda < 1 $, we expect the transition to the Y-x  insulating phase \cite{rh}  is driven by a roton touching
%at $ (\pi, 0 ) $  from both PW-X  SF at $ 0 < \beta < \beta_c $ and  PW-XY SF at $ \beta > \beta_c $.
%At $ \lambda=1 $, the transition from $ Y-x $ SF to the  $ Y-x $ state is in the same universality class of
%conventional SF to Mott transition.

\begin{table*}[ht]
\caption{
The quantum information scrambling  behaviours of the pseudo-Goldstone mode and the slow-Goldstone mode achieved by the non-perturbative approach developed here.
	$\Delta_{R}^{-}\sim \sqrt{1-\lambda}$, $\Delta_{R}^{+}\sim \lambda-1 $.
The massive scalar at $ \lambda > 1 $ can be reached just by perturbation theory and is listed here only for completeness and comparison.
%The ratio of the velocity of the  resurgent SOC Goldstone mode over that of the SF Goldstone mode is $ r_G=v_{soc}/c \sim 1 \% $.   
}
\renewcommand{\arraystretch}{2}
\setlength{\tabcolsep}{5pt}
%\begin{ruledtabular}
%\begin{tabular}{c|c|c|c|c|c}
\begin{tabular}{cccccc}	
    \toprule
    Deformation &  Particle & Spurious  & Corrected  &  Lyapunov exponent & Butterfly velocity \\
\hline
	$ \lambda < 1 $  & pseudo-Goldstone &  $ \omega \sim k $ 	&   $ \omega \sim  \sqrt{ (\Delta_{R}^{-})^2 + v^2 k^2 } $  &   $ \lambda_L\sim e^{-\Delta_{R}^{-}/T}  $    & 
  $ v_B \sim v \sqrt{ T/\Delta_{R}^{-} },~~ T \ll \Delta_{R}^{-}  $   \\
	$ \lambda = 1 $  &   slow-Goldstone      & $ \omega \sim k^2  $   &  $ \omega \sim v_{soc} k  $  &  $ \lambda_L\sim T^{3}/\rho^2_{s,R} $   &  $ v_B\sim v_{soc} \ll c  $     \\
%\hline \hline
    $ \lambda > 1 $  & massive scalar & ---   &  $ \omega \sim \Delta_{R}^{+}  + t k^2 $  &  $  \lambda_L\sim e^{-\Delta_{R}^{+}/T} $   &  $ v_B \sim \sqrt{ T t}, ~~ T \ll \Delta_{R}^{+} $   \\  
\hline
\end{tabular}
\label{tab}
%\end{ruledtabular}
\end{table*}

Taking two initially commuting operators separated by a large spatial distance $ x $,
$[W(x,0),V(0,0)]=0$,
and assuming $ \langle  V \rangle_{\beta}=  \langle  W \rangle_{\beta} =0 $, 
one can study how the commutator evolves over time under the Hamiltonian $ H $:
\begin{equation}
  C(x, t)=  \langle  [ W(x, t),  V(0,0)]^{\dagger}_{\pm} [ W(x, t),  V(0,0)]_{\pm} \rangle_{\beta}\>,
\end{equation}
where $ W(x,t)= e^{i H t} W (x,0)e^{-i H t} $ and the $ \pm $ applies
  when $ W $ and $ V $ are fermionic  and bosonic  operators respectively.
  The  average $ \langle \cdots \rangle_{\beta} $ could be the infinite temperature $ \beta=0 $ or  any finite temperature.
  
  For a quantum chaotic system, the quantum information scrambling encoded  in $ C(x,t) $ can be described by:
 \begin{equation}
    C( x, t)  = f_0 -f_1 e^{ \lambda_L(t- |x |/v_B) }
\label{lcone}
\end{equation}
    where $ \lambda_L  $ is the Lyapunov exponent, $ v_B $ is the butterfly velocity.
    The physical picture is the following:
under a quantum chaotic evolution, the operator $ W(x,t) $ will grow ballistically with a characteristic quantum butterfly velocity $ v_B $.
The $ W(x,t) $ and $ V(0,0) $ still commute, so $ C(x,t) $ remains small, until a scrambling time $ t_s > |x|/v_B $ when $ V(0,0) $ enters the
butterfly light cone of $ W(x,t) $. So it is  the ballistic growth of the commutators in the spatial directions which defines the butterfly light-cone $ |x|=v_B t $ around which
the quantum information scrambling increases exponentially with the Lyapunov exponent $ \lambda_L $ in  Eq.\eqref{lcone}.
The butterfly velocity $ v_B $, in general, should satisfy the Lieb-Robinson bound on the commutator
of local operators separated in time for systems with local interactions \cite{LR}, so $ v_B $ may also be called Lieb-Robinson velocity.  
In the present context, $ W= V $ can be chosen as the density or spin 
operator of the system. Then it represents the quantum information scramblings in the density and spin channel respectively.

{\sl (1) Enhanced Quantum information scramblings due to the %resurgent 
	SOC Goldstone mode }

One usually classify quantum chaos as the following 5 classes in terms of the Lyapunov exponent $ \lambda_L $ at a low temperature $ T $:
(i) maximal quantum chaos $ \lambda_L= 2 \pi T/\hbar $, 
(ii) quantum chaos  $ \lambda_L \sim  aT  $ where $ a < 2 \pi/\hbar $,
(iii) weak  quantum chaos  $ \lambda_L \sim  T^{b} $ where $ b > 1  $,
(iv)  suppressed quantum chaos $ \lambda_L \sim  e^{-T/\Delta }  $ where $ \Delta $ is the gap of the system, and finally 
(v)  an integrable system  $ \lambda_L = 0  $.

 It is the gap which determines the Lyapunov exponent $ \lambda_L $, while it is the dispersion which determines the butterfly velocity $ v_B $ 
 with the corresponding light cone $ x-v_B t $ (Table \ref{tab}).
Because the superfluid (or density) sector is non-critical, 
so we only focus on the roton (or SOC) sector in Fig.\ref{rotondrive}.
%(see also Appendix ).
It is important to point out that the "QPT" driven by the spin-anisotropy interaction $ \lambda $ in  Fig.\ref{rotondrive} is 
different than the conventional one: there is an exact $ U(1)_{soc} $ symmetry at $ \lambda=1 $, so in contrast to the enlarged emergent symmetry 
due to the spontaneous symmetry breaking where there is an enhanced quantum chaos  $ \lambda_L \sim  aT  $ where $ a < 2 \pi/\hbar $
(for example, the quantum Lifshitz transition driven by the superfluid Goldstone boson at $ \beta=\beta_c $ in Fig.\ref{sfhalf} to be mentioned in the conclusion),
here it is the exact symmetry, so it is just has  a weak  quantum chaos  $ \lambda_{L,R} \sim  T^{3}/\rho^2_{s,R}$ where  $ \rho_{s,R} $
is the superfluid density in the SOC sector which, 
as computed in the Appendix B, 
is much smaller than that in the superfluid sector, 
so the coefficient in the front of $ T^3 $ becomes quite large. 
The corresponding butterfly velocity   $ v_{B,R} =v_{soc}  \sim \sqrt{n_0} U  $ listed below Eq.\eqref{rotonl}  is also very tiny as shown in Fig.\ref{rotondrive}(b).
When $ \lambda \neq 1 $, a gap is opened with $ \Delta_R \sim |1- \lambda |^{\nu } $ where $  \nu=1/2, 1  $ listed in Eq.\eqref{gap} and Eq.\eqref{gap2} respectively, so 
the quantum chaos  is exponentially suppressed with $ \lambda_{L,R} \sim  e^{-T/\Delta_R }  $.
The corresponding butterfly velocity is given by $  v_{B,Rx}=\sqrt{ \frac{B_R}{\Delta_R} T } $ if $ T \ll \Delta_R $ or 
 $  v_{B,Rx}=\sqrt{ B_R } $  if $ T > \Delta_R $ in Eq.\eqref{gapdisp} and $ V_{B,Rx}= \sqrt{ t T }   $ if $ T \ll \Delta_R $  in Eq.\eqref{lambdalarge}.
These facts may also be considered as the enhancement of quantum chaos  at the 
``QPT'' driven by the spin-anisotropy interaction $ \lambda $ in  Fig.\ref{rotondrive}(a).
%In the light  $ \lambda $ dependence and $ n_0 U $ dependence.

%\noindent
%{\bf  Detections on cold atom experiments with Rashba SOC in optical lattices: }
\subsection{Detections in cold-atom experiments}

The pesudo-spin-1/2 boson
can be experimentally realized using 
ultracold akaline atoms. %such as $^{23}$Na, $^{41}$K, or $^{87}$Rb.
For example, in $^{87}$Rb ($5S_{1/2}$), 
nuclear spin $I=3/2$ leads to an $F=1$ electronic ground state.
One can choose
with hyper-fine state 
$|\!\!\uparrow\rangle=|F=1,m_F=-1\rangle$
and
$|\!\!\downarrow\rangle=|F=1,m_F=0\rangle$,
while the third state 
$|F=1,m_F=+1\rangle$ 
can be removed by a sufficiently large two-photon detuning\cite{2dsocbec}.
As suggested in \cite{uniform3}, 
SOC may be realized in near future experiments 
by adding spin-flip Raman lasers or 
by driving the spin-flip transition with radio‑frequency or microwave fields. 
%However, heating issues associated with spontaneous emissions 
%need to be overcame to study many-body phenomena.
%Recently, a new experiment using Raman schemes successfully realized a 2d Rashba SOC in $^{40}$K Fermi gas \cite{expk40,expk40zeeman}
%However, it is still not known if many body effects can be observed in this experiment set-up.
Due to the heating issues associated with the SOC generated 
by Raman laser scheme on alkali fermions, 
it would be difficult to observe many body phenomena on alkali fermions
with the Raman scheme \cite{expk40,expk40zeeman}.
However, optical lattice clock schemes \cite{clock} have been successfully implemented 
to generate SOC for $^{87}$Sr \cite{clock2} and $^{173}$Yb \cite{clock1}. 
This newly developed scheme has the advantage to suppress
the heating issue suffered in the Raman scheme.
It can also probe the interplay between the interactions and the SOC easily.
% The PI believe that it is just a matter of time for the experimentalists worldwide to realize various kinds of SOC \cite{nodoubt}.

%  So in principle, all the parameters in Eqn.\eqref{intlambda} $ t $, $ U >0 $ or $ U<0 $ and the SOC parameters
%  $ \alpha, \beta, \gamma $ can be tuned independently in a cold atom experiment.
%  A tremendous advantages of the 3d cubic lattice systems discussed in IV over the 2d square and honeycomb lattices in I-III is
%  that it supports much higher $ T_c $.

   The time of flight (TOF) image after a time $ t $ is given by \cite{blochrmp}:
\begin{equation}
	 n( \mathbf{r} )= ( M/\hbar t )^3 f(\mathbf{k}) G(\mathbf{k})\>,
\end{equation}
where $ \mathbf{k}= M \mathbf{r}/\hbar t $,  
$ f(\mathbf{k})= | w(\mathbf{r}) |^2 $ is the form factor due to the
   Wannier state of the lowest Bloch band of the optical lattice and
   $ G(\mathbf{k}) = \frac{1}{N_s} \sum_{i,j} e^{- \mathbf{k} \cdot ( \mathbf{r}_i- \mathbf{r}_j ) }
    \langle \Psi^{\dagger}_i \Psi_j \rangle $ is the equal time boson structure factor.
   For small condensate depletion in the weak interaction limit $ U/t \ll 1 $,
   $ \langle \Psi^{\dagger}_i \Psi_j \rangle \sim \langle \Psi^{\dagger}_{0i} \Psi_{0j}  \rangle$
   where $ \Psi^{\dagger}_{0i} $ is the condensate wavefunction Eq.\eqref{twonodes} at the mean field level.
   So the TOF can detect the quantum ground state wavefunction such as the PW-X and Z-x state in Fig.\ref{twoexp} directly.

The anisotropic interaction $ \lambda $ in Eq.\eqref{intlambda} can be easily tuned between two species of spinor bosons in two different hyperfine states.
 All the superfluid phases, their excitation spectrum and 
 the transition driven by a tiny gapped pseudo-Goldstone mode touch down at $ \lambda < 1 $ to 
 become the soft %resurgent 
 SOC Goldstone mode at $ \lambda=1 $
 , then a large gapped pseudo-Goldstone mode at $ \lambda > 1 $ shown in 
 Fig.\ref{twoexp} can be precisely determined by various experimental techniques
 such as dynamic or elastic, energy or momentum resolved, longitudinal or
 transverseBragg spectroscopies \cite{braggbog-1,braggbog-2,braggbog-3,braggbog-4,braggbog-5},
 specific heat measurements \cite{heat1,heat2}  and {\sl In-Situ} measurements \cite{dosexp}.
 The very narrow butterfly light-cone  $ x-v_{B,R}t $ along the time direction  due to the %resurgent 
 Goldstone boson in the SOC sector 
%(Fig.\ref{rotondrive}(a)) 
 can also be detected by the techniques developed in \cite{OTOCexp1,OTOCexp2}.
% In fact, the compressibility $ \kappa $ Ref.\cite{heat2} and the superfluid density $ \rho_s $ \cite{sfdensity} of the Goldstone mode
% can be separately measured.

%\noindent
\section{DISCUSSION}

%\noindent
 There are two kinds of Goldstone modes in the SOC interacting system studied here.
 One is the superfluid Goldstone mode due to the $ U(1)_c $ symmetry breaking. Its velocity can be easily calculated just 
 by the conventional perturbative method.
 Another is the slow-Goldstone mode  at $ \lambda=1 $ due to the $ U(1)_{soc} $ breaking.
 Its tiny velocity can only be computed by the new non-perturbative formalism developed here.
 The two Goldstone modes play very different roles, especially in displaying quantum information scramblings  in terms of Lyapunov exponent and 
 butterfly velocity (see Table \ref{tab}).
 This slow-Goldstone may be contrasted to the slow light  \cite{slowlight} which
 is the propagation of an optical pulse or other modulation of an optical carrier at a very low group velocity. 
 Of course, complete different mechanisms are responsible for the slowness of the spin-0 and spin-1 bosons respectively.
 The former is due to the spurious symmetry at the classical level and the non-perturbative OFQD mechanism developed in this work.  
 The latter  occurs when a propagating pulse is substantially slowed by the interaction with the medium in which the propagation takes place.
 
% But both involve interactions. It remains interesting to see how to use the Resurgent SOC Goldstone

 It is also  interesting to  contrast the pseudo-Goldstone and Goldstone mode discovered here by the %resurgent 
 formalism here with the soft Goldstone mode
 identified by $1/N $ expansion in the Sachdev-Ye-Kitaev (SYK) models.
 In the SYK models \cite{SY,kittalk,syk2,on}, the saddle point solution spontaneously  breaks the reparametrization symmetry to the global $ SL(2, Z) $ symmetry,
resulting a zero Goldstone mode. Then when considering the leading irrelevant operator $ -i \omega $ perturbatively  in the $ 1/N $ expansion which also
breaks the reparametrization symmetry explicitly,  it will lift the zero Goldstone mode to a gapless (soft)  Pseudo-Goldstone mode described by the   Schwartizian.
which leads to the maximal quantum chaos. So in the SYK case, a leading order in $ 1/N $ expansion is enough to capture the main physics of SYK models: the maximal chaos which is dual to that of a quantum black hole. 
There is no need for any non-perturbative formalism 
to capture the soft pseudo-Goldstone mode.  Due to the lack of the space dimension in the SYK model, there is no dispersion relation for such a soft mode, also no associated butterfly velocity. 
As stressed in the Sec.V, in the present interacting SOC system, there are the intrinsic connections between the quantum chaotic properties of the system and those of the elementary excitations as listed in the Table \ref{tab}. 
It is the slow-Goldstone mode at $ \lambda =1 $ which leads to the enhanced quantum chaos characterized by the power law Lyapunov exponent and a very slow butterfly velocity.  It is also the small gap $ \Delta_{R}^{-} $ of the pseudo-Goldstone mode at $ \lambda < 1 $ which leads to the suppressed quantum chaos characterized by the exponentially suppressed Lyapunov exponent and a power law vanishing butterfly velocity at  a low temperature $ T \ll \Delta_{R}^{-} $.

Here we develop a new and systematic non-perturbative approach to investigate this new class of QPT induced by the roton touchdown at $ \lambda =1 $ 
  leading to  the %resurgent 
  slow-Goldstone mode. The superfluid Goldstone mode remains un-critical through such a process.
  The quantum Lifshitz transition induced by the superfluid Goldstone mode at $ \beta=\beta_c $ in Fig.\ref{sfhalf}  will be addressed elsewhere.
As shown in \cite{rh}, at integer fillings and $ \lambda=1 $, 
the system described by Eq.\eqref{intlambda} 
enters the Mott insulating Y-x phase in the strong coupling limit
($U/t\gg 1$).
We note that the Y-x superfluid phase shown in Fig. \ref{blochsphere},
which occurs in the weak coupling limit ($U/t \ll 1$), 
shares the same spin-orbital structure as this Mott insulating phase. 
It is therefore interesting to investigate potential 
quantum and topological phase transitions between these states. 
Such transitions may occur through intermediate quantum spin liquid phases 
as increasing $U/t$, connecting the Y-x superfluid phase 
(with both superfluid Goldstone and slow-Goldstone bosons) to the Y-x Mott phase.
The non-perturbative approach developed here 
could be widely used to study novel emergent quantum phenomena 
in many other frustrated systems,  
also in particle physics related to the 
Coleman-Weinberg effective potential \cite{CWpotential}.
%and possibly in some branches of Mathematics such as 
%%dynamical system, partial diferential equations and Geometry.
%%and may inspire new approaches in
%dynamical systems, partial differential equations, and geometric analysis.
Additionally,
our formalism might exhibit superficial similarities to the
resurgent theory developed in topological strings and matrix models,
though establishing any rigorous connections requires further investigation.

\section*{Acknowledgments}

We thank Jing Zhou for helpful discussions.  
This work is supported by Guangdong Basic and Applied Basic Research Foundation (Grant No. 2024A1515010698).

\medskip

\textit{Note added.---} After we finished this work, a new experiment \cite{newsu2} appeared, it realized a tunable $ SU(2) $ gauge field of spinor bosons
of $^{87}$Rb atoms loaded in a one dimensional ladder system.

%The Rashba SOC (QAH) model is topological trivial ( non-trivial ).
%So this work is complementary to that  in the QAH model studied in \cite{SFQAH}.
%        The results to be achieved may shed lights on each other.
%        It  possible deep connections between
%the many body phenomena of weakly interacting spinor bosons and the topology of the underlying non-interacting energy bands are also explored recently..
%        It is constructive to compare Fig.\ref{Rashba}b with  Fig.\ref{review1}c.
%        The spin anisotropy $ \lambda-1 $ and the SOC parameter $ \beta $ in Fig.\eqref{Rashba}b play similar roles as the Zeeman field $ h $
%        and $ t_s/t $ in Fig.\eqref{review1}c.
%        Let us focus on the $ \lambda=1 $ in Fig.\eqref{Rashba}b and $ h=0 $ in Fig.\eqref{review1}c.
%        The Y-x SF ( with two minima ) and IC-SF ( with four minima ) at $ \lambda=1 $
%        in the former keeps and breaks the $ U(1)_{soc} $ symmetry respectively, while the
%        C-SF ( with two minima ) and the IC-SF ( with four minima ) in the latter keeps and breaks the $ Z_2 $ symmetry at $ H=0 $ respectively.
%     Especially, similar " order from quantum disorder" analysis was employed to study various SF in both models.
%       can also be employed to study  the Y-x SF in Fig.\eqref{Rashba}b.
%``rrttt"

%  F. Sun and J. Y acknowledge AFOSR FA9550-16-1-0412 for the supports at earlier stage of this work.
%  The work at KITP was supported by NSF PHY11-25915.

%\onecolumngrid
%\clearpage

%\medskip
%\noindent
%{\bf METHODS }

\appendix

%\noindent
%{\bf The Pseudo-Goldstone mode and resurgent Goldstone mode generated by the Resurgent mechanism:  resurgent Canonical Quantization approach }
\section{Canonical quantization approach to %the resurgent mechanism}
capture the non-perturbative effect}

In this appendix, a unified non-perturbative scheme is developed
to compute the whole excitation spectrum, 
extending well beyond the gap-calculation scheme in \cite{gan} 
and containing it at the zero momentum point. 
This approach automatically includes not only 
the mass generation of the pseudo-Goldstone mode at $\lambda<1$, 
but also the computation of the spectrum of the slow-Goldstone mode at $\lambda=1$.

We perform our non-perturbative analysis using the Z-x representation 
Eq.~\eqref{Zxbasis}, 
though equivalent results may be obtained in the dual Y-x representation 
Eq.~\eqref{Yxbasis},
which is related to the Z-x basis by the duality transformation in Eq.~\eqref{exch}
and is employed in Sec. IV (see Fig.~\ref{blochsphere}).
%So we may just use the OFQD  variables $\phi$ introduced in the $ \lambda < 1 $ case.
Since $B(\lambda)$ in Eq.\eqref{A2B2} and Eq.\eqref{B2} 
is continuous at $\lambda=1$ as shown in Fig.\ref{Bterm1}(a), 
we only need focus on the $\lambda<1$ case
and then take the $ \lambda \rightarrow 1^{-} $ limit.
We need to establish the relation between the OFQD variables $\phi$ in Eq.\eqref{A2B2} 
in the old OFQD analysis and the original Bose fields in Eq.\eqref{intlambda}.

The Bose condensation Eq.\eqref{twonodes} inspires us to parameterize
the most general Bose field in the polar-like coordinate system as
\begin{align}
    \Psi_i
	=e^{i\chi}\sqrt{\rho}
	\left[
	c_1\eta_1
	e^{i\mathbf{K}\cdot\mathbf{r}_i}
	+
	c_2\eta_2
	e^{-i\mathbf{K}\cdot\mathbf{r}_i}
	\right]\>,
\label{bosonoriginal}
\end{align}
where one can parameterize the two coefficients $c_{1,2}$
in the most general forms in the Z-x representation in Eq.\eqref{c1c2}
%\begin{align}
%	c_1&=[e^{i\phi/2}\cos(\theta/2)+e^{-i\phi/2}\sin(\theta/2)]/\sqrt{2},  \nonumber  \\
%	c_2&=[e^{i\phi/2}\cos(\theta/2)-e^{-i\phi/2}\sin(\theta/2)]/\sqrt{2},
%\label{c1c2Zx}
%\end{align}
and the two-component spinors $\eta_{1,2}$ as:
\begin{align}
	\eta_n=
	\begin{pmatrix}
	    e^{+i\phi_n/2}\cos(\theta_n/2)\\
	    e^{-i\phi_n/2}\sin(\theta_n/2)\\
	\end{pmatrix},
	~	
	n=1,2.
\label{spinorsQF}
\end{align}
Thus we can write 
$\Psi_i=\Psi_i(\rho,\chi,\theta,\phi,\theta_1,\phi_1,\theta_2,\phi_2)$, 
which also takes into account the quantum fluctuations 
of the two spinors $\eta_{1,2}$ .

Since we already found the quantum ground state is the PW-X SF state
which corresponds to the saddle point values:
\begin{align}
    \Psi_{i,0} =\Psi_{i}(\rho_0,0,\pi/2,0,-\pi/2,0,\pi/2,0)\>.
\label{saddlePWX}
\end{align}
One can write down the quantum fluctuations around the saddle point as
$\Psi_i=\Psi_{i,0} + \delta\Psi_i$,
where $\delta\Psi_i$ can be expressed as a function of
$\delta\rho,\chi,\delta\theta,\phi,
\delta\theta_1,\phi_1,\delta\theta_2,\phi_2$.
%\begin{align}
%$\Psi_i=\Psi_{i,0} + \Psi_i( \delta\rho,\chi,\delta\theta,\phi,
%\delta\theta_1,\phi_1,\delta\theta_2,\phi_2 ) $.
%\end{align}
Thus, we can separate the Bose field into the condensation part 
plus the quantum fluctuation part in the polar coordinate system:
\begin{align}
    \Psi_i
	& =\sqrt{\frac{\rho_0}{2}}
	\begin{pmatrix}
	    1\\
	   -1\\
	\end{pmatrix}
	e^{i\mathbf{K}\cdot\mathbf{r}_i}    \nonumber  \\
	& +
	\sqrt{\frac{\rho_0}{8}}
	\left[
	(\frac{\delta\rho}{\rho_0}+i\chi)
	\begin{pmatrix}
	    1\\
	    -1\\
	\end{pmatrix}
	%e^{i\mathbf{K}\cdot\mathbf{r}_i}
	+
	%\sqrt{\frac{\rho_0}{8}}
	(\delta\theta_1+i\phi_1)
	\begin{pmatrix}
	    1\\
	    1\\
	\end{pmatrix}
	\right]
	e^{i\mathbf{K}\cdot\mathbf{r}_i}     \nonumber    \\
   & + \sqrt{\frac{\rho_0}{8}}
	(-\delta\theta+i\phi)
	\begin{pmatrix}
	    1\\
	    1\\
	\end{pmatrix}
	e^{-i\mathbf{K}\cdot\mathbf{r}_i}\>,
\label{poldecomp}
\end{align}
where the third line explicitly contains the OFQD variable $ \phi $.

On the other hand,
in the original Cartesian coordinate, we have:
\begin{align}
    \Psi_i=\sqrt{\frac{N_0}{2}}
	\begin{pmatrix}
	    1\\
	   -1\\
	\end{pmatrix}
	e^{i\mathbf{K}\cdot\mathbf{r}_i}
	+
	\begin{pmatrix}
	    \psi_{i\uparrow}\\
	    \psi_{i\downarrow}\\
	\end{pmatrix}\>,
\label{cardecomp}
\end{align}
After comparing both the $e^{i\mathbf{K}\cdot\mathbf{r}_i}$ 
and the $e^{-i\mathbf{K}\cdot\mathbf{r}_i}$
components in the two equations Eq.\eqref{poldecomp} and Eq.\eqref{cardecomp}, 
one can express the OFQD variable $\phi$ in terms of the original Bose field.
After a Fourier transformation this leads to
\begin{align}
    \phi_\mathbf{q}
     =\frac{-i}{\sqrt{2\rho_0}}
	[\psi_{-\mathbf{K}+\mathbf{q},\uparrow}
	+\psi_{-\mathbf{K}+\mathbf{q},\downarrow}   %\nonumber  \\
	-\psi^{\dagger}_{\mathbf{K}-\mathbf{q},\uparrow}
	-\psi^{\dagger}_{\mathbf{K}-\mathbf{q},\downarrow}] \>,
\end{align}
where, as expected, 
only the quantum fluctuations near $ -\mathbf{K} $ appear in the relation.
Thus we can express the quantum correction coming from the OFQD mechanism 
in terms of the original Bose fields as:
%Quantum order-from-disorder mechanism lead to a correction to Hamiltonian
\begin{align}
	\delta \mathcal{H}
	=\sum_i\frac{B}{2}\phi_i^2
	=\sum_q\frac{B}{2}\phi_q\phi_{-q}\>,
\label{corr}
\end{align}
which, we must stress, only holds in the momentum regime near $ -\mathbf{K} $, 
so it will not affect the Goldstone mode near $ \mathbf{K} $.

Finally, after combining with $\mathcal{H}^{(2)}$ in Eq.\eqref{h2} 
(its expression before applying the Bogoliubov transformation, see Appendix C),
we arrive at the non-pertubative Hamiltonian:
\begin{align}
	\mathcal{H}
	=\mathcal{H}^{(2)}+\delta \mathcal{H}
	=\frac{1}{2}\sum_{q}\Psi_q^\dagger (M+\delta M) \Psi_q\>,
\label{total}
\end{align}
where the $4\times4$ matrix $M$ was already obtained from 
the PW-X SF calculation leading to Eq.\eqref{h2},
and the matrix $\delta M$ %in Eq.\eqref{corr} 
can be written in a $ 4 \times 4 $ matrix form:
\begin{align}
    \delta M=
	\frac{B}{2n_0}
	\begin{pmatrix}
		1 &1 &-1 &-1\\
		1 &1 &-1 &-1\\
		-1&-1&1  &1\\
		-1&-1&1  &1\\
	\end{pmatrix}\>.
\end{align}
 We diagonalize the Hamiltonian Eq.\eqref{total}
 by a $ 4 \times 4 $ Bogoliubov transformation:
\begin{align}
    \mathcal{H}
	=\sum^{2}_{l=1} \sum_{q \in BZ }\omega_{l}(\mathbf{q})
	\left(\beta_{l,q}^\dagger\beta_{l,q}+\frac{1}{2}\right)\>,
\end{align}
where $\omega_{1}(\mathbf{q})\leq\omega_{2}(\mathbf{q})$. 
Thus one only needs to focus on the lowest band $ \omega_1 $. 
Note that here we started from the PW-X SF state in Eq.\eqref{saddlePWX}, 
so $\mathbf{q}$ is defined in the whole Brillouin zone, 
in contrast to the OFQD analysis Eq.\eqref{h2ground} 
where $\mathbf{k}$ defines in the reduced Brillouin zone.

From the complete form of the dispersion 
$ \omega_1 ( -\mathbf{K} + \mathbf{q} ) $,
we can extract the long wavelength limit of 
the Pseudo-Goldstone mode as listed in Eq.\eqref{gapdisp}
\begin{align}
    \omega_R(\mathbf{q})
	= \sqrt{\Delta^2_R + B_R
	 [ q_x^2 +(\cos\beta-C\sin\beta^2 ) q_y^2]}\>,
\end{align}
where
$\Delta_R=\omega_R(\mathbf{q}=0)=\sqrt{2BU(1-\lambda)}$ 
is the roton gap in Eq.\eqref{gap}
generated by the 2nd step OFQD mechanism already developed in \cite{gan}. 
So our unified scheme recovers the previously developed OFQD 
at the zero momentum, but goes much further.
The two coefficients $B_R=2Bt/n_0+n_0Ut(1-\lambda)$ and
$ C=1+\frac{BU[4t(1-3\lambda)+2B(1-\lambda)/n_0+n_0U(1-\lambda)^2]}
{[2B/n_0+n_0U(1-\lambda)][8t^2+2n_0Ut(1+\lambda)-BU(1-\lambda)]}<1 $
represent the corrections to the roton dispersion,
which can only be achieved by the new non-perturbative fromalism developed here.
These terms are important and lead to the butterfly velocity of the pseudo-Goldstone mode:
$v_{B,Rx}=\sqrt{ (B_R/\Delta_R) T } $ if $ T \ll \Delta_R $ 
or  $v_{B,Rx}=\sqrt{B_R}$  if $ T > \Delta_R $.
At the quantum cricial point $ \lambda=1 $, 
it becomes the butterfly velocity $v_{soc}$ of 
the slow-Goldstone boson in Eq.\eqref{rotonl}.
Obviously, if one sets $ B=0 $ by mistake, 
Eq.\eqref{gapdisp} recovers the second equation in Eq.\eqref{GR}, 
which is the linear spurious roton mode.

Of course, due to the momentum separation as stressed below Eq.\eqref{corr},
the superfluid Goldstone mode $ \omega_1 ( \mathbf{q} ) $ 
dictated by the $ U_c(1) $ symmetry breaking
remains the same as listed the first equation in Eq.\eqref{GR}.

It is important to stress that the Eq.\eqref{gapdisp} 
goes well beyond Eq.\eqref{gap} in many respects:  
it not only contains the roton gap $\Delta_R $ 
in Eq.\eqref{gap} at the  zero momentum $ q=0 $ ,
but also the roton dispersion relation at long wavelength 
due to the non-pertubative effect.
This is the first non-pertubative calculation of the correction 
to the dispersion relation beyond just a gap calculation in \cite{gan}.
Furthermore, by taking $\lambda \to 1^-$ limit in Eq.\eqref{gapdisp},
the roton gap vanishes and recovers the linear dispersion 
in Eq.\eqref{rotonl} for the slow-Goldstone mode.
So it is the non-pertubative formalism which leads to both 
the Lyapunov exponent $\lambda_{B,R}$, the butterfly velocity $v_{B,R}$,
and the associated light cone  $x-v_{B,R}t$ 
in both the pseudo-Goldstone mode at $ \lambda < 1 $ 
and the slow-Goldstone mode at $ \lambda=1 $.

%\noindent
%{\bf The Pseudo-Goldstone mode and resurgent Goldstone mode generated by the  Resurgent  mechanism:  the Resurgent Path integral approach }
\section{Path integral approach to 
capture the non-pertubative effect}

We perform our non-perturbative analysis in the Z-x representation,
though equivalent results may be obtained in the Y-x representation 
(Fig.\ref{blochsphere}).
The path integral approach is complementary 
to the canonical approach presented in Appendix A.

In the weak interaction limit, 
the low-energy physics can be captured by 
phase-amplitude representation of original boson 
in Eq.\eqref{bosonoriginal}:
\begin{align}
    b(\tau)\!=\!\!\sqrt{n(\tau)}e^{i\chi(\tau)}
	[c_1(\tau)\eta_1e^{i\mathbf{K}_1\cdot\mathbf{r}_i}
	\!+\!c_2(\tau)\eta_2e^{i\mathbf{K}_2\cdot\mathbf{r}_i}],
\end{align}
where $c_{1,2}$ are given in the Z-x representation Eq.\eqref{c1c2}. 
Here, we focus on the two lowest excitation modes: 
superfluid (density) mode and the roton (SOC) mode, 
so we drop the gapped quantum fluctuations 
in the two spinors $\eta_{1,2}$ in Eq.\eqref{spinorsQF}.

The Lagrangian density takes the form
\begin{align}
	\mathcal{L}
		&=\sum_{k} b_k^\dagger \partial_\tau b_k+\mathcal{H}[b,b^\dagger] \nonumber\\
	&\approx
	\sum_{k} b_k^\dagger (\partial_\tau+\epsilon_{k}^{-}-\mu) b_k
		-E_\text{int}[\{c_i\}]\>,
\end{align}
where the mean-field interaction energy density takes the form Eq.\eqref{classic}
\begin{align}
    E^{0}_\text{int}
		%=\frac{Un^2}{4}\left[1+\lambda+(1-\lambda)\cos^2\theta\right]
		=\frac{Un^2}{2}\left[1+\frac{\lambda-1}{2}\sin^2\theta\right]\>,
\label{classic1}
\end{align}
which determines $\theta=\pi/2$ but cannot determine $\phi$.
However, the OFQD analysis indicates that the PW-X SF is the ground state,
which means $(c_1,c_2)=(1,0)$ or $(c_1,c_2)=(0,1)$,
thus suggests $\phi=0$.
Plugging in $\theta=\pi/2$ gives $c_1=\cos(\phi/2)$ and $c_2=i\sin(\phi/2)$.

The quantum ground-state (saddle point solution) corresponds to
$(n,\chi,\theta,\phi)=(n_0,0,\pi/2,0)$
and the quantum fluctuations can be written 
as the deviations around the saddle point solution:
$n=n_0+\delta n,~ \chi=0+\delta\chi,~
\theta=\pi/2+\delta\theta,~ \phi=0+\delta\phi $.
Working out fluctuations in the original Hamiltonian
and dropping unimportant constants yields the Lagrangian:
\begin{align}
    \mathcal{L}=&i\delta n\partial_\tau\chi
	+i\frac{n_0}{2}\delta\theta\partial_\tau\delta\phi\nonumber\\
	&+\frac{1}{4}tn_0[\frac{1}{n_0^2}(\tilde{\nabla}\delta n)^2
		+4(\tilde{\nabla}\delta\chi)^2
		+(\tilde{\nabla}\delta\theta)^2
		+(\tilde{\nabla}\delta\phi)^2]	\nonumber\\
	&+\frac{1}{4}U[(1+\lambda)(\delta n)^2+(1-\lambda)n_0^2(\delta\theta)^2]\>,
\label{L-SF}
\end{align}
where $\tilde{\nabla}=(\partial_x,\sqrt{\cos\beta-\sin^2\beta}\partial_y)$.

Besides, we can expand the quantum ground-state energy as
\begin{align}
	\delta E(c_1,c_2)=\frac{1}{2}B(\delta\phi)^2
\label{correct}
\end{align}
%where $B$ .
By adding it to the Lagrangian to reach a new Lagrangian, 
one can solve for its two eigen modes 
$ (\delta n, \delta \chi) $ and  $ (\delta \theta, \delta \phi) $
in the momentum space:
%\begin{align}
%	\omega_1
%		=&\sqrt{\big[tk_x^2+t(\cos\beta-\sin^2\beta)k_y^2+(1+\lambda)n_0U\big]
%			   \big[tk_x^2+t(\cos\beta-\sin^2\beta)k_y^2\big]}\\
%	\omega_2
%		=&\sqrt{\big[tk_x^2+t(\cos\beta-\sin^2\beta)k_y^2+(1-\lambda)n_0U\big]
%			   \big[tk_x^2+t(\cos\beta-\sin^2\beta)k_y^2+2B/n_0\big]}
%\end{align}
\begin{align}
    \omega_1=&\sqrt{g_\mathbf{k}[g_\mathbf{k}+(1+\lambda)n_0U]} \>, \nonumber\\
    \omega_2=&\sqrt{[g_\mathbf{k}+(1-\lambda)n_0U][g_\mathbf{k}+2B/n_0]} \>,
\end{align}
where $g_\mathbf{k}=tk_x^2+tu(\beta)k_y^2$.
As expected, the superfluid Goldstone mode 
$\omega_1$ is independent of the $ B $ term.

The non-perturbative analysis generates a gap 
$\Delta_R=\sqrt{2BU(1-\lambda)}$ 
for the pseudo-Goldstone mode $\omega_2$ at $\mathbf{k}=0$, 
but extends significantly to the full spectrum.
Since $B$ remains finite as $\lambda\to1$, 
the roton gap behaves as
$\Delta_R\propto\sqrt{1-\lambda}$ for $\lambda<1$. 
Results from both complementary approaches are consistent. 
Dropping the $B$ term and taking the long-wave length limit
recovers $\omega_1$ and $\omega_2$ from Eq.~\eqref{GR}.

In fact, one can extract the superfluid and roton mode separately 
from Eq.\eqref{L-SF} and Eq.\eqref{correct}, one can integrate out 
$\delta n$ and $\delta\phi$ which are conjugate to  $ \delta \chi $ and $ \delta \theta $ in Eq.\eqref{L-SF}, and obtain
\begin{align}
	\mathcal{L}_\text{eff}
	=&\frac{1}{U(1+\lambda)}(\partial_\tau\chi)^2+n_0t(\tilde{\nabla}\delta\chi)^2
	+\frac{n_0^2}{8B}(\partial_\tau\delta\theta)^2 \nonumber\\
	&+\frac{1}{4}n_0t(\tilde{\nabla}\delta\theta)^2
	+\frac{1}{4}(1-\lambda)n_0^2U(\delta\theta)^2\>.
\end{align}

After taking into account the anisotropy encoded in the derivative 
$\tilde{\nabla}=(\partial_x,\sqrt{\cos\beta-\sin^2\beta}\partial_y)$,
one can read out the zero-temperature 
compressibility and superfluid density 
from the density channel $ \delta \chi $:
\begin{align}
	\chi^\text{SF}= \frac{1}{U(1+\lambda)},~~
	\rho_{x}^\text{SF}=n_0t,~~
	\rho_{y}^\text{SF}=n_0t u(\beta) \>,%(\cos\beta-\sin^2\beta),
\end{align}
and also the extra slow-Goldstone mode at $\lambda=1$ 
from the SOC channel $ \delta \theta $:
\begin{align}
	\chi^\text{soc}=\frac{n_0^2}{8B},~~
	\rho_{x}^\text{soc}=\frac{1}{4}n_0t,~~
	\rho_{y}^\text{soc}=\frac{1}{4}n_0tu(\beta) \>. %(\cos\beta-\sin^2\beta).
\end{align}
When $\beta$ is small, the anisotropy becomes small. 
Note that  $\chi^\text{soc} \ll \chi^\text{SF} $ 
leads to the relation 
$ v_{soc}=\sqrt{\rho^\text{SF}/\chi^\text{SF}} 
\ll  c=\sqrt{\rho^\text{soc}/\chi^\text{soc}}$.

At low temperatures, the superfluid densities in both sectors decrease.
They vanish via universal jumps at critical temperatures: 
$T_\text{KT}^\text{SF}$ for the charge $U(1)_c$ symmetry breaking 
and $T_\text{KT}^\text{soc}$ for the spin-orbit $U(1)_{\text{soc}}$ symmetry breaking. 
Following Landau's approach \cite{Fetter}, 
the superfluid density follows 
$\rho_{s}(T)=\rho_{s}(0)-\rho_{n}(T)$, 
where the normal density 
$\rho_n(T)\propto T^3/v^4$ depends on the velocity:
$v = c$ for the superfluid Goldstone mode 
and $v = v_{\text{soc}}$ for the slow-Goldstone mode.
The small ratio $r_G$ between the slow-Goldstone and superfluid Goldstone mode velocities 
yields the hierarchy $T_\text{KT}^\text{soc} \ll T_\text{KT}^\text{SF}$, 
as shown in Fig. \ref{rotondrive}(a).

\begin{widetext}

\section{Details on the series expansion Eq.\eqref{Hasym}   }

%\medskip

%\noindent
In order to unify calculations for all $\lambda$ case,
we take the most general state 
and rewrite the Bose field into a mean-field part plus a fluctuating part
\begin{align}
    \begin{pmatrix}
	b_{\mathbf{k}\uparrow}	\\
	b_{\mathbf{k}\downarrow}\\
    \end{pmatrix}
    =
    \sqrt{N_0}\left[
    \frac{c_1}{\sqrt{2}}
    \begin{pmatrix}
	1\\
	-1\\
    \end{pmatrix}
    \delta_{\mathbf{k},\mathbf{K}}
    +
    \frac{c_2}{\sqrt{2}}
    \begin{pmatrix}
	1\\
	1\\
    \end{pmatrix}
    \delta_{\mathbf{k},-\mathbf{K}}
    \right]
    +
    \begin{pmatrix}
	\psi_{\mathbf{k}\uparrow}\\
	\psi_{\mathbf{k}\downarrow}\\
    \end{pmatrix}\>,
\end{align}
where $c_1$ and $c_2$ are two complex numbers 
subject to the normalization condition $|c_1|^2+|c_2|^2=1$,
and $N_0$ is the number of condensate atoms.
Performing the above substitution,
lead to an expansion of original Hamiltonian 
\begin{align}
    \mathcal{H}=\mathcal{H}^{(0)}+\mathcal{H}^{(1)}+\mathcal{H}^{(2)}+\cdots\>,
\end{align}
where the superscript denotes the order in the fluctuations
and $\cdots$ means high order.
After some algebras, 
the $\mathcal{H}^{(0)}$ term takes the form %classic mean-field energy
\begin{align}
    \mathcal{H}^{(0)}
	=-n_0[2t(1+\cos\beta)+\mu]
	+\frac{Un_0^2}{2}\Big(1+\frac{\lambda-1}{2}[1-(c_1c_2^*+c_1^*c_2)^2]\Big)\>,
\end{align}
and $\mathcal{H}^{(1)}$ term can be easily worked out
\begin{align}
	\mathcal{H}^{(1)}=&
	\sqrt{\frac{N_0}{2}}
	\Big(
	[-2t(1+\cos\beta)-\mu]c_1
	+\frac{Un_0}{2}[(1+\lambda)c_1+(1-\lambda)c_2(c_1c_2^*+c_1^*c_2)]
	\Big)
	(\psi_{\mathbf{K}\uparrow}^\dagger-\psi_{\mathbf{K}\downarrow}^\dagger) \nonumber\\
	&+\sqrt{\frac{N_0}{2}}
	\Big(
	[-2t(1+\cos\beta)-\mu]c_2
	+\frac{Un_0}{2}[(1+\lambda)c_2+(1-\lambda)c_1(c_1c_2^*+c_1^*c_2)]
	\Big)
	(\psi_{-\mathbf{K}\uparrow}^\dagger+\psi_{-\mathbf{K}\downarrow}^\dagger)+h.c. \>.
\end{align}
From $\mathcal{H}^{(1)}=0$, 
the chemical potential can be determined, along with a constraint on $c_{1,2}$:
\begin{alignat}{3}
    &\text{when }\lambda<1,\quad 
	&&\mu=-2t(1+\cos\beta)+ Un_0(1+\lambda)/2,\quad   
	    &c_1c_2^*+c_1^*c_2=0;    \label{case1}\\
    &\text{when }\lambda=1,\quad 
	&&\mu=-2t(1+\cos\beta)+ Un_0,\quad   
	    &\text{any }c_1c_2^*+c_1^*c_2;   \label{case2}\\
    &\text{when }\lambda>1,\quad
	&&\mu=-2t(1+\cos\beta)+ Un_0,\quad
	    &c_1c_2^*+c_1^*c_2=\pm1;  \label{case3}
\end{alignat}
then $\mathcal{H}^{(2)}$ term is a quadratic theory
\begin{align}
    \mathcal{H}^{(2)}
	&=\sum_\mathbf{k} \psi_\mathbf{k}^\dagger h_k \psi_\mathbf{k}
	-\mu\sum_\mathbf{k} \psi_\mathbf{k}^\dagger\psi_\mathbf{k}
	+\frac{Un_0}{2}
	\sum_\mathbf{k}[
	(2+\lambda)(\psi_{\mathbf{k}\uparrow}^\dagger\psi_{\mathbf{k}\uparrow}
		+\psi_{\mathbf{k}\downarrow}^\dagger\psi_{\mathbf{k}\downarrow})
	+\lambda(|c_2|^2-|c_1|^2)
	      (\psi_{\mathbf{k}\uparrow}^\dagger\psi_{\mathbf{k}\downarrow}
	      +\psi_{\mathbf{k}\downarrow}^\dagger\psi_{\mathbf{k}\uparrow})] \nonumber\\ 
	&+\frac{Un_0}{4}
	\sum_\mathbf{k}[
	2c_1c_2
	(\psi_{\mathbf{k}\uparrow}^\dagger\psi_{-\mathbf{k}\uparrow}^\dagger
	-\psi_{\mathbf{k}\downarrow}^\dagger\psi_{-\mathbf{k}\downarrow}^\dagger)
	+(c_1^2+c_2^2)
	(\psi_{\mathbf{k}\uparrow}^\dagger\psi_{-\mathbf{k}+\mathbf{Q}\uparrow}^\dagger
	+\psi_{\mathbf{k}\downarrow}^\dagger\psi_{-\mathbf{k}+\mathbf{Q}\downarrow}^\dagger)
	+2\lambda(c_2^2-c_1^2)\psi_{\mathbf{k}\uparrow}^\dagger\psi_{-\mathbf{k}+\mathbf{Q}\downarrow}^\dagger
	+h.c.] \nonumber\\
	&+\frac{Un_0}{2}
	\sum_\mathbf{k}[(2-\lambda)(c_1c_2^*+c_1^*c_2)
	      (\psi_{\mathbf{k}\uparrow}^\dagger\psi_{\mathbf{k}+\mathbf{Q}\uparrow}
	      -\psi_{\mathbf{k}\downarrow}^\dagger\psi_{\mathbf{k}+\mathbf{Q}\downarrow})
	+\lambda(c_1c_2^*-c_1^*c_2)
	      (\psi_{\mathbf{k}\uparrow}^\dagger\psi_{\mathbf{k}+\mathbf{Q}\downarrow}
	      -\psi_{\mathbf{k}\downarrow}^\dagger\psi_{\mathbf{k}+\mathbf{Q}\uparrow})] \>,
\label{eq:H(2)}
\end{align}
where we define $\mathbf{Q}=2\mathbf{K}=(\pi,0)$.
To incorporate both spin and valley degrees of freedom, a spinor is introduced as:
$\Psi_\mathbf{k}=
(\psi_{\mathbf{K}+\mathbf{k},\uparrow},
\psi_{\mathbf{K}+\mathbf{k},\downarrow},
\psi_{-\mathbf{K}+\mathbf{k},\uparrow},
\psi_{-\mathbf{K}+\mathbf{k},\downarrow})^\intercal$.
In this notation, the Hamiltonian $\mathcal{H}^{(2)}$ can be expressed in matrix form as
\begin{align}
    \mathcal{H}^{(2)}=
	\frac{1}{2}\sum_{\mathbf{k}\in\text{RBZ}}\text{Tr}M_{22}(\mathbf{k})
	+
	\frac{1}{2}\sum_{\mathbf{k}\in\text{RBZ}}
	\begin{pmatrix}
	    \Psi_\mathbf{k}^\dagger  &\Psi_{-\mathbf{k}}^\intercal
	\end{pmatrix}
	\begin{pmatrix}
	    M_{11}(\mathbf{k}) &M_{12}(\mathbf{k})\\
	    M_{21}(\mathbf{k}) &M_{22}(\mathbf{k})\\
	\end{pmatrix}
	\begin{pmatrix}
	    \Psi_\mathbf{k}\\
	    \Psi_{-\mathbf{k}}^{\dagger\intercal}\\
	\end{pmatrix}\>,
\label{eq:H(2)M}
\end{align}
where $M_{ij}$ for $i,j=1,2$ are $4\times4$ matrix,
and the overall $8\times8$ matrix is referred to as $M$.
%Diagonal part $M_{11}$ and $M_{22}$ are describing hopping,
%off-diagonal part $M_{12}$ and $M_{21}$ are describing pairing.
Note the off-block-diagonal part $M_{12}$ and $M_{21}$ 
are comes from anomalous terms in Eq.\eqref{eq:H(2)}.
The bosonic commutation relation and Hermiticity yield the relationships
$M_{11}(\mathbf{k})=M_{22}(-\mathbf{k})^\intercal$ 
and $M_{12}(\mathbf{k})=[M_{21}(\mathbf{k})]^\dagger$.
It is convienient to decomposite $M$ into three parts:
$M=M^{(t)}+M^{(\mu)}+M^{(U)}$,
where $M_{ij}^{(\mu)}=\mu\delta_{ij}I_{4\times4}$,
$M_{12}^{(t)}=M_{21}^{(t)}=0$ and 
\begin{align}
	M_{11}^{(t)}(\mathbf{k})\!=\!-2t
	\begin{pmatrix}
		\cos\beta\cos k_y & -\cos k_x+i\sin\beta\sin k_y & 0 &0\\
		-\cos k_x-i\sin\beta\sin k_y &\cos\beta\cos k_y &0 &0\\
		0 &0 &\cos\beta\cos k_y &\cos k_x+i\sin\beta\sin k_y\\
		0 &0 &\cos k_x-i\sin\beta\sin k_y &\cos\beta\cos k_y\\
	\end{pmatrix},
\end{align}
and the block-diagonal part of $M^{(U)}$
\begin{align}
	M_{11}^{(U)}(\mathbf{k})\!=\!Un_0\!
    \begin{pmatrix}
	1+\lambda/2 &(\lambda/2)(|c_2|^2-|c_1|^2) &(1-\lambda/2)(c_1c_2^*+c_1^*c_2) &(\lambda/2)(c_1c_2^*-c_1^*c_2)\\
	(\lambda/2)(|c_2|^2-|c_1|^2) &1+\lambda/2 &\lambda/2(c_1^*c_2-c_1c_2^*) &(\lambda/2\!-\!1)(c_1c_2^*+c_1^*c_2)\\
	(1\!-\!\lambda)(c_1c_2^*+c_1^*c_2) &(\lambda/2)(c_1c_2^*-c_1^*c_2) &1+\lambda/2 &(\lambda/2)(|c_2|^2-|c_1|^2)\\
	(\lambda/2)(c_1^*c_2-c_1c_2^*) &(\lambda/2-1)(c_1c_2^*+c_1^*c_2) &(\lambda/2)(|c_2|^2-|c_1|^2) &1+\lambda/2
    \end{pmatrix},
\end{align}
as well as off-block-diagonal part of $M^{(U)}$
\begin{align}
    M_{12}^{(U)}(k)=Un_0
    \begin{pmatrix}
	(c_1^2+c_2^2)/2 &(\lambda/2)(c_2^2-c_1^2) &c_1c_2 &0\\
	(\lambda/2)(c_2^2-c_1^2) &(c_1^2+c_2^2)/2 &0 &-c_1c_2\\
	c_1c_2  &0 &(c_1^2+c_2^2)/2 &(\lambda/2)(c_2^2-c_1^2)\\
	0 &-c_1c_2 &(\lambda/2)(c_2^2-c_1^2) &(c_1^2+c_2^2)/2\\
    \end{pmatrix}\>.
\end{align}
It is well known that
Hamiltonian Eq.\eqref{eq:H(2)M} can be diagonlized by a $8\times8$ Bogoliubov transformation
\begin{align}
	&\mathcal{H}^{(2)}=
	\frac{1}{2}\sum_{\mathbf{k}\in\text{RBZ}}
	\text{Tr}M_{22}(\mathbf{k})
	+
	\frac{1}{2}\sum_{\mathbf{k}\in\text{RBZ}}
	\begin{pmatrix}
	    \upbeta_\mathbf{k}^\dagger  
	    &\upbeta_{-\mathbf{k}}^\intercal
	\end{pmatrix}
	\begin{pmatrix}
	    \Omega(\mathbf{k}) &0\\
	    0 &\Omega(-\mathbf{k})\\
	\end{pmatrix}
	\begin{pmatrix}
	    \upbeta_\mathbf{k}\\
	    \upbeta_{-\mathbf{k}}^{\dagger\intercal}\\
	\end{pmatrix}\\
    &\upbeta_\mathbf{k}
	=\mathbf{u}_\mathbf{k}\Psi_\mathbf{k}
	 +\mathbf{v}_{-\mathbf{k}}\Psi_{-\mathbf{k}}^{\dagger\intercal},
    \qquad
    \upbeta_\mathbf{-k}^{\dagger\intercal}
	=\mathbf{u}_{-\mathbf{k}}^*\Psi_{-\mathbf{k}}^{\dagger\intercal}
	 +\mathbf{v}_\mathbf{k}^*\Psi_\mathbf{k},
\end{align}
where
$\Omega(\mathbf{k})=\text{diag}
(\omega_{1,\mathbf{k}},\omega_{2,\mathbf{k}},
\omega_{3,\mathbf{k}},\omega_{4,\mathbf{k}})$,
$\upbeta_\mathbf{k}=(
\beta_{1,\mathbf{k}},
\beta_{2,\mathbf{k}},
\beta_{3,\mathbf{k}},
\beta_{4,\mathbf{k}})$
and $\mathbf{u}_\mathbf{k}$, $\mathbf{v}_\mathbf{k}$ are two $4\times4$ matrix.
To ensure the Bogoliubov transformation is canonical,
$\beta_{l,\mathbf{k}}$ must satisfy bosonic commutation relation,
i.e.
$[\beta_{l,\mathbf{k}},\beta_{l^\prime,\mathbf{k}'}^\dagger]
=\delta_{l,l'}\delta_{\mathbf{k},\mathbf{k}'}$,
and $[\beta_{l,\mathbf{k}},\beta_{l',\mathbf{k}'}]=0$.
These conditions will give constraint to $\mathbf{u}_\mathbf{k}$ and $\mathbf{v}_\mathbf{k}$,
which results in a secular equation
\begin{align}
	\det
	\begin{pmatrix}
	M_{11}(\mathbf{k})-\omega_{l,\mathbf{k}}I &M_{12}(\mathbf{k})\\
	M_{21}(\mathbf{k}) &M_{22}(\mathbf{k})+\omega_{l,\mathbf{k}}I\\
	\end{pmatrix}
	=0
\label{eq:secularEq}
\end{align}
where $I$ stand for a $4\times4$ identity matrix,
and $\omega_l$ are eigen-energy after Bogoliubov transformation.

After combined condition Eq.(\ref{case1}-\ref{case3})
and the secular equation Eq.\eqref{eq:secularEq}
and then solve it,
we accomplished the diagonalization of the quadratic theory 
\begin{align}
	\mathcal{H}^{(2)}=E_0^{(2)}+
	 \sum_{\mathbf{k}}
	 \sum_{l=1}^{4}\omega_{l,\mathbf{k}}
	 \left(\beta_{l,\mathbf{k}}^\dagger\beta_{l,\mathbf{k}}+\frac{1}{2}\right)
\label{eq:H2bog}
\end{align}
In general, it is challenging to diagonalize the matrix analytically; 
however, numerical methods can always be employed to extract useful information.

%\medskip
%\noindent
%{\bf The evaluation of the velocity ratio of the Resurgent  SOC Goldstone mode over the SF Goldstone mode   }
\section{Details on the evaluation of the velocity ratio $v_\text{soc}/c$ }

This section provides technical details on the evaluation of the velocity 
of the slow-Goldstone mode. 
The following calculations are performed under the condition $\lambda\leq1$,
and the parameterization of $c_1$ and $c_2$ from Eq.(6) is used.

%The key point is that the PW-X SF solution ($\phi = 0$) can make 
%the secular equation Eq.\eqref{eq:secularEq} simply enough to solve analytically.

For generic $\phi$, the 4 Bogoliubov bands $\omega_{l}(\mathbf{k}; \phi)$ 
%in Eq.~\eqref{h2ground}
in Eq.~\eqref{eq:H2bog}
are quite complicated, but they simplify considerably at $\phi = 0$ as:
\begin{align}
	\omega_{1}(\mathbf{k}; \phi=0) = \sqrt{A_1+\sqrt{A_1^2-B_1^2}},\quad
	\omega_{2}(\mathbf{k}; \phi=0) = \sqrt{A_1-\sqrt{A_1^2-B_1^2}},\\ \nonumber 
	\omega_{3}(\mathbf{k}; \phi=0) = \sqrt{A_2+\sqrt{A_2^2-B_2^2}},\quad
	\omega_{4}(\mathbf{k}; \phi=0) = \sqrt{A_2-\sqrt{A_2^2-B_2^2}},
\label{Bigfour}
\end{align}
where we define
\begin{align}
	A_s&=h_0(h_0+n_0U)+h_x^2+h_y^2+(-1)^s n_0U\lambda h_x,\\ \nonumber 
	B_s^2&=(h_0^2-h_x^2-h_y^2)
	[h_0^2-h_x^2-h_y^2+n_0U(2h_0-2\lambda h_x+(1-\lambda^2)n_0U)],\\   \nonumber 
	h_0&=2t(1+\cos\beta)-2t\cos\beta\cos k_y,
	\quad 
	h_x=2t\cos k_x,
	\quad
	h_y=-i2t\sin\beta\sin k_y\>.
\end{align}
However, the secular equation of arbitrary $\phi$
can be significantly simplified using the results obtained at $\phi=0$:
\begin{align}
	0%\text{Eq.(S15)}
	=(\omega^2-\omega_1^2)(\omega^2-\omega_2^2)
	 (\omega^2-\omega_3^2)(\omega^2-\omega_4^2)
	 -f\>,
\label{gse}
\end{align}
where $\omega_i=\omega_{i}(\mathbf{k}; \phi=0)$ 
and $f=f(\omega,\phi)$ with $f(\omega,\phi=0)=0$.
Nevertheless, it is generally an eighth-degree equation, 
and there is no analytical method to obtain an exact solution,
although numerical methods are always available.

The ground-state energy $E_\text{GS}(\phi)$ in Eq.~\eqref{h2ground} is
\begin{align}
	E_\text{GS}(\phi)=E_{0t}
	    +\frac{1}{2}\sum_{l,\mathbf{k}}\omega_{l}(\mathbf{k}; \phi),
\end{align}
which is ploted as a function of $\phi$ for various $\lambda$ in 
Fig.\ref{Bterm1}(a).
The numerical data of $E_\text{GS}(\phi)$ shows minima at $\phi=0$ and $\phi=\pi$.
%So we only need 
In order to evaluate $B$ analytically, 
one need take derivative with respect to $\phi$ at $\phi=0,\pi$
\begin{align}
	B(n_0U,t,\beta,\lambda)
	=\frac{\partial^2}{\partial \phi^2}\Big|_{\phi=0}
	\frac{1}{2}
	\sum_{\mathbf{k}\in \text{RBZ}}\sum_{l=1}^{4}\omega_{l}(\mathbf{k};\phi)\>.
\end{align}
Thus derivative of $\omega_{l}(\mathbf{k};\phi)$ with respect to $\phi$
is crucial in determining $B$.
The perturbative approach to solving the general secular equation
Eq.\eqref{gse} yields these derivatives.
Thus, the exact analytical expression of $B$ is obtained as
\begin{align}
	B(n_0U,t,\beta,\lambda)
	=
	&\sum_\mathbf{k}
	\frac{g_1(\omega_1+\omega_2+\omega_3+\omega_4)}
	{(\omega_1+\omega_2)(\omega_1+\omega_3)(\omega_1+\omega_4)
	 (\omega_2+\omega_3)(\omega_2+\omega_4)(\omega_3+\omega_4)} 
	\nonumber\\
	&+\sum_\mathbf{k}
	\frac{g_2(\omega_1\omega_2\omega_3+\omega_1\omega_2\omega_4
			+\omega_1\omega_3\omega_4+\omega_2\omega_3\omega_4)}
	{(\omega_1+\omega_2)(\omega_1+\omega_3)(\omega_1+\omega_4)
	 (\omega_2+\omega_3)(\omega_2+\omega_4)(\omega_3+\omega_4)}
	 \nonumber\\
	&+\sum_\mathbf{k}
	\frac{g_3[(\omega_1+\omega_2+\omega_3+\omega_4)^3
		-\omega_1^3-\omega_2^3-\omega_3^3-\omega_4^3]}
	{3\omega_1\omega_2\omega_3\omega_4
	 (\omega_1+\omega_2)(\omega_1+\omega_3)(\omega_1+\omega_4)
	 (\omega_2+\omega_3)(\omega_2+\omega_4)(\omega_3+\omega_4)}\>,
\end{align}
where  $\omega_l\equiv\omega_l(k;\phi=0)$ are listed in Eq.\eqref{Bigfour}  and 
\begin{align}
	g_1&=16n_0^2U^2h_0^2h_y^2-4(1-\lambda^2)n_0^2U^2h_y^2(3h_0^2-h_x^2-h_y^2+n_0Uh_0)\>,\\
	g_2&=2(1-\lambda^2)n_0^2U^2h_y^2\>,\\
	g_3&=2(1-\lambda^2)n_0^2U^2h_y^2(h_0^2-h_x^2-h_y^2)
	[h_0^2-h_x^2-h_y^2+2n_0Uh_0+(1-\lambda^2)n_0^2U^2]\>.
\end{align}

At $\lambda=1$, $B$ takes the simpler form:
\begin{align}
	B(n_0U,t,\beta,\lambda=1)
	=\sum_\mathbf{k}
	\frac{16n_0^2U^2h_0^2h_y^2(\omega_1+\omega_2+\omega_3+\omega_4)}
	{(\omega_1+\omega_2)(\omega_1+\omega_3)(\omega_1+\omega_4)
	(\omega_2+\omega_3)(\omega_2+\omega_4)(\omega_3+\omega_4)}\>.
\end{align}
It is easy to scale out the overall factor $(n_0U\sin\beta)^2/t$ and 
rewrite $B=b (n_0U\sin\beta)^2/t$ with a dimensionless quantity $b$.
The limiting value of $b$ as $\beta\to0$ and $U\to0$ can be analytical worked out as
\begin{align}
	\lim_{\beta\to0}\lim_{U\to0}b
	=
	\lim_{\beta\to0}\lim_{U\to0}
	\frac{B(\lambda=1)}{(n_0U\sin\beta)^2/t}
	=\frac{3-\sqrt{3}\ln(2+\sqrt{3})}{12\pi} \approx 0.01907\>.
\end{align}
The $B$ and $b$ are ploted as a function of $n_0U/t$ for various $\beta$ 
in Fig.\ref{Bterm1}(b).%
At $\lambda=1$, the ratio between 
the slope of slow-Goldstone mode and that of the superfluid Goldstone mode is given by
\begin{align}
	r_G
	=\frac{\sqrt{2Bt/n_0}}{\sqrt{2n_0tU}}=\sqrt{\frac{B}{n_0^2U}}
	=\sqrt{\frac{b(n_0U)^2\sin^2\beta/t}{n_0^2U}}
	=\sqrt{b(U/t)\sin^2\beta}\>.
\label{eq:r-G}
\end{align}
The typical value for $r_G$ is $\approx 1\%$ shown in Fig.\ref{rotondrive}(b).

\end{widetext}


\begin{thebibliography}{99}
\bibitem{notation} 
Here we still use the same notation used in \cite{rh}.
The PW-X means the plane wave state with the spin polarization along the X direction.
The PW-XY means the plane wave state with the spin polarization along the XY direction.
In the Z-x phase, the first Capital letter indicates
the spin polarization, the second small letter indicates the orbital order.
\bibitem{subtract}  
Because the bosons can condense at one of the $ N_K=2 $ minima in Fig.1b. 
In the following, the momentum is measured relative to this condensation point.
\bibitem{inter} 
In this manuscript, we use the roton or pseudo-Goldstone mode interchangeably.
\bibitem{NK=4}
For $\beta>\beta_c$, 
the two minima split into four minima at incommensurate momenta $(\pm\pi/2,\pm k_0)$. 
The physics of this $N_K=4$ phase when $\beta_c<\beta<\pi/2$, 
including the quantum Lifshitz transition at $\beta=\beta_c$ 
induced by the superfluid Goldstone boson, 
will be discussed in a separate publication.
\end{thebibliography}
\end{document}